\documentclass[aps,physrev,reprint,superscriptaddress]{revtex4-2}

\usepackage{graphicx}
\usepackage{orcidlink}
\usepackage{physics}
\usepackage{soul}
\usepackage{xcolor}
\usepackage{float}

\begin{document}
\title{Room-temperature inversionless diamond nitrogen-vacancy electronic spin maser}
\author{\orcidlink{0009-0007-7498-0741}Ali Fawaz}
\affiliation{School of Mathematical and Physical Sciences, Macquarie University, North Ryde, NSW 2109, Australia}
\author{\orcidlink{0000-0002-9081-5750}Sarath Raman Nair}
\email[Corresponding author]{}
\affiliation{School of Mathematical and Physical Sciences, Macquarie University, North Ryde, NSW 2109, Australia}
\date{\today}
\begin{abstract}
We propose a method to create a room-temperature maser operating at approximately 2.9~GHz frequency using an ensemble of negatively charged nitrogen-vacancy electronic spins (NV) in diamond, without requiring population inversion. 
Our method considers a DC magnetic field of a few milli-Tesla (mT) applied along the perpendicular direction of an ensemble of NV spins aligned along a common axis.
This perpendicular magnetic-field creates superposition states of $|m_{\mathrm{s}}=-1\rangle$ and $|m_{\mathrm{s}}=+1\rangle$ of the NV spin's ground state triplet levels and thereby makes it possible to drive all three transitions in the NV spin ground state.
We model the system by including optical pumping of the NV spins, near-resonant driving of two transitions, and coupling the third transition to a near-resonant microwave resonator.
Numerical estimates using experimentally realizable parameters show that inversionless masing can be achieved inside the microwave resonator using our method.
As an application, we show that the output intensity of an inversionless maser ($1.1\times10^{14}$ spins) can be used for magnetic field sensing with a DC sensitivity on the order of a hundred pT/$\sqrt{\mathrm{Hz}}$.
Our study opens a new direction in room-temperature diamond NV maser devices for quantum technological applications without the requirement of a strong bias magnetic field, as in conventional NV diamond masers.
\end{abstract}
\maketitle
\paragraph*{Introduction \textemdash}The room-temperature continuous-wave (CW) maser using an optically pumped negatively charged nitrogen-vacancy electronic (NV) spin ensemble in diamond \cite{jin2015proposal, breeze2018continuous, zollitsch2023maser, sherman2022diamond, day2024room, ng2025portable} offers a platform for quantum technological applications \cite{arroo2021perspective}.
For example, a diamond NV maser can be used as a magnetic-field sensor \cite{jin2015proposal, arroo2021perspective}.
On the other hand, the conventional diamond NV maser \cite{breeze2018continuous, zollitsch2023maser, sherman2022diamond, day2024room, ng2025portable, jin2015proposal, arroo2021perspective} requires a large magnetic field, typically provided by a heavy and bulky magnet, which limits the practical applicability.
The state-of-the-art diamond maser device in terms of compactness and lightness is the one reported in \cite{ng2025portable}, to the best of our knowledge, which uses a microwave-oven-sized electromagnet with a mass of approximately 30~kg to generate a 400~mT magnetic field for a maser frequency of around 9.6~GHz. 
Though this magnetic field requirement can be reduced if the maser frequency is smaller, as the diamond NV maser frequency is tunable via magnetic-field, for example, a 3~GHz maser requires around 210~mT \cite{jin2015proposal}, the required field remains substantial.
In this work, we propose a method to create a room-temperature inversionless maser \cite{harris1989lasers, mompart2000lasing} in CW mode and at a frequency around 2.9~GHz, using diamond NV spins.
We show that this maser can be used for magnetic field sensing via changes in the output intensity induced by the target magnetic field, similar to the room-temperature p-terphenyl pentacene maser magnetic field sensor \cite{wu2022enhanced} or the proposed diamond room-temperature NV laser magnetometer \cite{jeske2016laser}.
Inversionless masing has been theoretically studied in a gain medium of diamond NV spins strongly hyperfine coupled to nearby nuclear spins \cite{wang2025cavity}.
Here, we focus on inversionless masing using NV spins alone, which is advantageous in the reproducibility of resultant technological devices. 

\begin{figure}
    \centering
    \includegraphics[width=\linewidth]{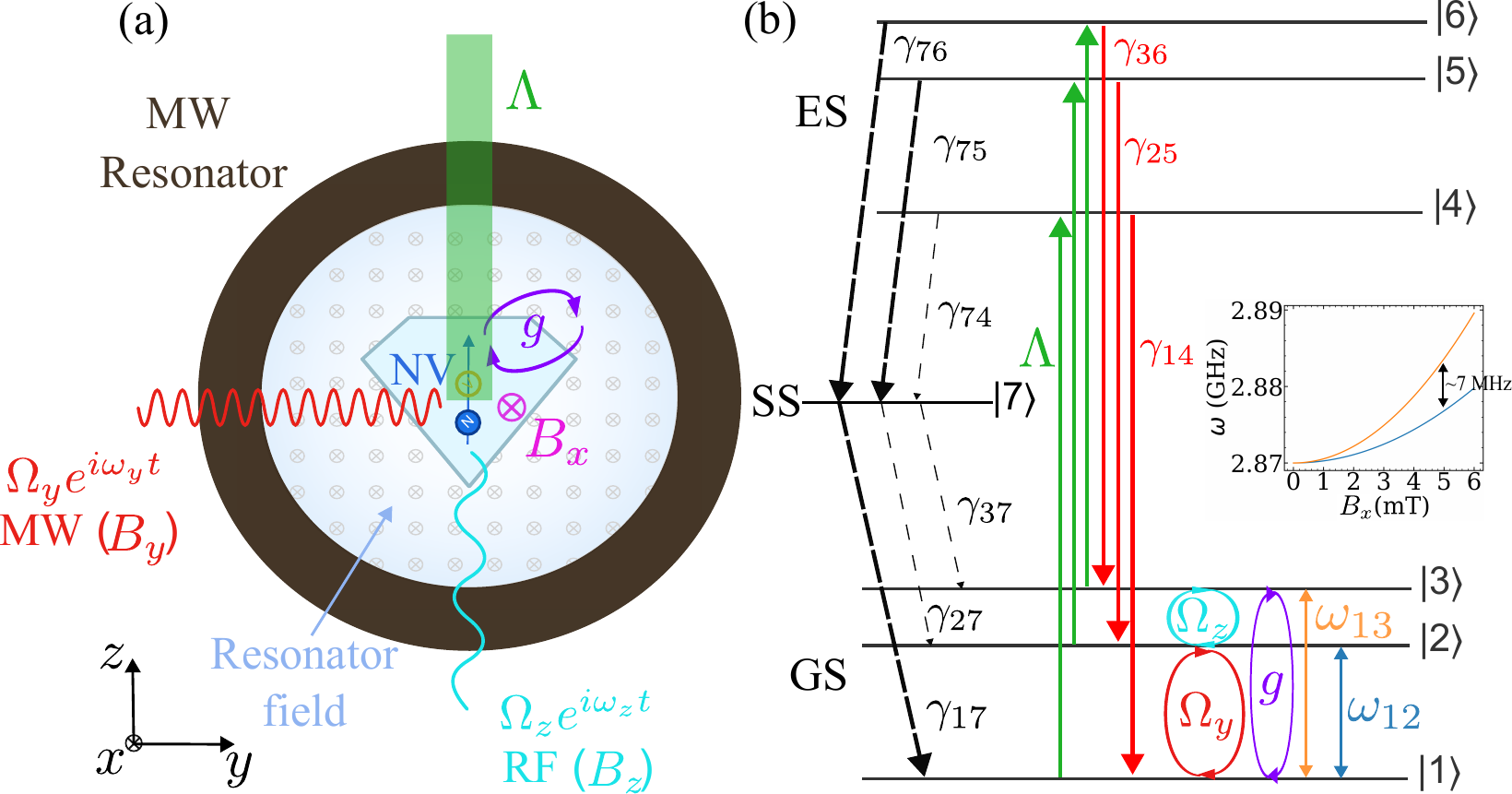}
    \caption{Proposed inversionless masing scheme. (a) Schematics of the inversionless diamond NV maser model. 
    A Cartesian coordinate system used in the model is defined by the NV spin axis ($z$-axis) and resonator B-field direction ($x$-axis).
    (b) Simplified seven-level scheme of NV spins used in our model, and the seven levels are labeled as $|1\rangle, |2\rangle,...$, and $|7\rangle$.
    All drivings on the ground-state transitions are indicated using the same color coding as in (a).
    The non-resonant optical excitation and direct radiative relaxation between the ground and excited states (GS and ES) are shown in green and red arrows.
    The simplified inter-system crossing (ISC) transitions from ES to GS through SS are shown in dashed arrows, where thicker ones have higher transition rates.
    The transition rates are denoted next to the respective transitions, and the suffix indicates the levels involved in the transition.
    The level splitting between $|2\rangle$ and $|3\rangle$ as a function of $B_{x}$ is also shown. 
    We consider $B_{x} = 5$~mT for our model, which corresponds to a frequency separation of 7~MHz between $|2\rangle$ and $|3\rangle$.}
    \label{fig:1}
\end{figure}
\setlength{\textfloatsep}{5pt}
\paragraph*{Model \textemdash}The proposed model system is presented schematically in Figure \ref{fig:1}.
We consider a diamond containing an ensemble of identical NV spins and assume a coordinate system with the NV spins oriented along the $z$-axis.
The ground state of the NV is a triplet state with magnetic quantum number $m_{s}=0, \pm1$, and we will denote the levels with $|..\rangle$ to represent the quantum state.
We consider that the NV spins are optically pumped using a green laser, which polarizes the spin population mostly to $|0\rangle$ via transitions involving the excited-state triplet and intermediate single states.
Unlike the conventional NV maser, we consider a bias magnetic-field, $B_{x}$, applied along the $x$-axis and assume that $B_{x} = 5~\text{mT}$, such that the field strength is small enough to induce any significant change in the optically induced spin dynamics of NV spins \cite{tetienne2012magnetic, lamba2024vector}.

The $B_x$ creates three new eigenstates $|0\rangle$, $(|+1\rangle-|-1\rangle)/\sqrt{2}$, and $(|+1\rangle+|-1\rangle)/\sqrt{2}$ \cite{matsuzaki2016optically, saijo2018ac, yamaguchi2019bandwidth, tabuchi2023temperature, mikawa2023electron}, since $\gamma_eB_x\ll D$ \cite{Supplementary_information}, where  $\gamma_e\,(\approx 28~\text{GHz/T})$ is the gyromagnetic ratio of the NV electronic spin, and $D \sim 2.87$~GHz \cite{gruber1997scanning} is the zero-field splitting. The resultant levels are denoted as $|1\rangle$, $|2\rangle$, and $|3\rangle$, respectively, for clarity in presentation.
Assuming $|1\rangle$ has zero reference energy, the transition frequencies in the new basis are $\omega_{12}=(D+\sqrt{D^2+(2\gamma_eB_x)^2})/2\approx 2.877~\text{GHz}$ and $\omega_{13}=\sqrt{D^2+(2\gamma_eB_x)^2}\approx 2.884~\text{GHz}$, where the suffixes represent the levels involved.
For simplicity, we do not consider strain-field effects, unlike in references \cite{matsuzaki2016optically, saijo2018ac, yamaguchi2019bandwidth, tabuchi2023temperature, mikawa2023electron, shin2013suppression}. This approximation is valid for strain fields with frequencies on the order of a few hundred kilohertz or lower.
All three resultant transitions $|1\rangle \leftrightarrow|3\rangle$, $|1\rangle \leftrightarrow|2\rangle$, and $|2\rangle \leftrightarrow|3\rangle$, have dipole moments along $x$-, $y$-, and $z$-axis, respectively \cite{lamba2024vector}.
We consider that the $|1\rangle \leftrightarrow |3\rangle$ transition of each spin is resonantly coupled to the resonator mode whose magnetic field is oriented along the $x$-axis. The $|1\rangle \leftrightarrow |2\rangle$ and $|2\rangle \leftrightarrow |3\rangle$ transitions are near-resonantly driven using additional microwave (MW) and radio-frequency (RF) fields, with the magnetic fields of the MW drive and RF drive oriented along the $y$-axis and $z$-axis, respectively.
The key idea of our method is to use the interference of spin coherences created using a MW and RF drive to suppress stimulated absorption in the $|1\rangle \leftrightarrow |3\rangle$, while still maintaining stimulated microwave photon emission to the resonator which is the underlying mechanism of lasing without population inversion \cite{harris1989lasers}, and thereby to achieve a maser output, whose intensity varies in the presence of an external magnetic field.

We denote the resonator frequency, the MW drive frequency, the RF drive frequency, single spin-resonator coupling strength, MW drive strength, and the RF drive strength as $\omega_{x}$, $\omega_{y}$, $\omega_{z}$, $g_{x}$, $\Omega_{y}$, and $\Omega_{z}$, respectively.
The total Hamiltonian for the interaction of the NV spin ground state in our system, in a suitable rotating frame with the rotating wave approximation, can then be written as \cite{Supplementary_information},
\begin{align}
\hat{H} = & \hbar \Delta_{x} \hat{a}^{\dagger}\hat{a} + \hbar \Delta_{12}  \sum_{k=1}^{N}\hat{\sigma}^{(k)}_{22} + \hbar \Delta_{13}  \sum_{k=1}^{N}\hat{\sigma}^{(k)}_{33}\nonumber\\ &+ \hbar g_{x}  \sum_{k=1}^{N}(\hat{a}^{\dagger} \hat{\sigma}^{(k)}_{13}+ \hat{a}  \hat{\sigma}^{(k)}_{31}) +i \hbar\Omega_{y} \sum_{k=1}^{N}(\hat{\sigma}^{(k)}_{12} -\hat{\sigma}^{(k)}_{21})\nonumber\\ &
+\hbar\Omega_{z}\sum_{k=1}^{N}( 
\hat{\sigma}^{(k)}_{23} + \hat{\sigma}^{(k)}_{32})).
\end{align}
Here, we use the spin operators $\hat{\sigma}^{(k)}_{\alpha\beta} = |\alpha\rangle \langle\beta|$ that describe the transition from $|\beta\rangle$ to $|\alpha\rangle$ for the $k$-th spin, and the photon creation (annihilation) operator $\hat{a}^\dagger$ ($\hat{a}$) of the resonator.
The parameters $\Delta_{x} = \omega_{x} -(\omega_{y}+\omega_{z})$, $\Delta_{12} = \omega_{12} -\omega_{y}$ and $\Delta_{13} = \omega_{13}-(\omega_{y}+\omega_{z})$, are detunings, and $N$ represent the number of spins.

To capture the effect of optical pumping,  we consider a seven-level model for the NV spins as shown in Figure \ref{fig:1} (b).
The optical excitation of NV spins with a green laser is an incoherent process \cite{doherty2011negatively, doherty2013nitrogen}, and we only consider coherence in the ground state in our model.
The optical excitation and direct radiative transitions between the ground and excited states (GS and ES) occur only between levels with the same $m_s$.
Since all the transitions to and from $m_{s} = + 1$ and $m_{s} = - 1$ are equal \cite{tetienne2012magnetic, gupta2016efficient}, we simplify our model with non-coherent transitions between $|2\rangle \leftrightarrow |5\rangle$, $|3\rangle \leftrightarrow |6\rangle$, and we do not expect any change in the dynamics of the system with this simplification.
We capture the effect of intersystem-crossing (ISC) transitions with only one long-lived singlet state (SS) (labeled as $|7\rangle$), similar to \cite{tetienne2012magnetic, gupta2016efficient} by neglecting the other singlet level whose lifetime is two orders of magnitude smaller than the one considered in \cite{rogers2008infrared, acosta2010optical, doherty2013nitrogen}.
We also simplify the ISC transitions in the same way as for the direct transitions between GS and ES.

We write down the equations of motion (EoM) for our system, similar to references \cite{zhang2022cavity, zhang2022microwave} using {\texttt{QuantumCumulants.jl}} \cite{plankensteiner2022quantumcumulants} in Julia \cite{Supplementary_information}, where for a general operator $\hat{O}$, the EoM is, $d\langle\hat{O}\rangle/dt = (i/\hbar)\langle\hat{H}\hat{O}-\hat{O}\hat{H}\rangle+\sum_{k} ( 
\hat{L}_k^{\dagger} \hat{O} \hat{L}_k - \frac{1}{2} \hat{L}_k^{\dagger} \hat{L}_k\hat{O} -\frac{1}{2} \hat{O} \hat{L}_k^{\dagger} \hat{L}_k)$.
Here $\hat{L}_{k}$ is the Lindbladian operator for losses in the system, and all $\hat{L}_{k}$ for our system are tabulated in \cite{Supplementary_information}.

We consider pure dephasing, longitudinal relaxation, and longitudinal excitation as losses for the ground state of the NV spins and in $|m_{\mathrm{s}}=0, \pm1\rangle$ basis they are $L_{p\pm} =\sqrt{2\Gamma}|\pm1\rangle\langle\pm1|$ \cite{zhang2022cavity, zhang2022microwave}, $L_{r\pm} = \sqrt{\gamma_{l}}|0\rangle\langle\pm1|$ and $L_{e\pm} =\sqrt{\gamma_{l}}|\pm1\rangle\langle0|$, respectively.
The parameters $\Gamma$ and $\gamma_{l}$ are the spin dephasing rate and longitudinal relaxation rate at room temperature, respectively.
To avoid underestimating the coherence loss in our model, we express them in the basis used in our model as $L_{p\pm}=\sqrt{(\Gamma/2)}(\hat{\sigma}^{(k)}_{22}+\hat{\sigma}^{(k)}_{33}\pm(\hat{\sigma}^{(k)}_{23}+\hat{\sigma}^{(k)}_{32}))$, $L_{r\pm} = \sqrt{(\gamma_{l}/2)}(\hat{\sigma}^{(k)}_{13}\pm\hat{\sigma}^{(k)}_{12})$, and $L_{e\pm} = \sqrt{(\gamma_{l}/2)}(\hat{\sigma}^{(k)}_{31}\pm\hat{\sigma}^{(k)}_{21})$.
Though $B_{x}$ could improve the $\Gamma$ if the primary source causing any NV spin dephasing is the electronic spin bath in the diamond \cite{shin2013suppression}, we neglect this effect for simplicity in the model.

The EoM for the mean photon-number of the resonator $n_x = \langle \hat{a}^{\dagger} a\rangle$ is $\dot{n}_x = (1+G) \kappa_{x}n_{\mathrm{th}} - \kappa_{x}n_{x}$, where $\kappa_{x}$ is the resonator photon loss rate, and $n_{\text{th}}=1/(\exp{\hbar\omega_x/k_BT}-1)$ is the thermal photon number given by the Bose-Einstein factor, where $k_B$ and $T$ are the Boltzmann factor and temperature, respectively.
The dimensionless parameter $G = - i(g_{x}/(\kappa_{x}n_{\mathrm{th}}))\sum_{k=1}^{N}(\langle \hat{a}^{\dagger} \hat{\sigma}^{k}_{13} \rangle - \langle \hat{a} \hat{\sigma}^{k}_{31} \rangle)$ characterizes gain, for example when $G>0$, we have a net gain in the resonator.
Assuming weak correlation between the operators, we can approximate $\langle \hat{a}^{\dagger} \hat{a} \rangle \approx \langle \hat{a}^{\dagger}\rangle\langle \hat{a} \rangle$, $\langle \hat{a}^{\dagger} \hat{\sigma}^{k}_{13} \rangle \approx \langle \hat{a}^{\dagger}\rangle\langle \hat{\sigma}^{k}_{13} \rangle$, and $\langle \hat{a} \hat{\sigma}^{k}_{31} \rangle \approx \langle \hat{a}\rangle\langle \hat{\sigma}^{k}_{31} \rangle$, and we can describe the system with a closed set of EoM in the first-order, which we write down as,
\begin{align}
\frac{d\langle \hat{a} \rangle}{dt} = &- i \sum_{k}^{N} g_{x} \langle \hat{\sigma}^{(k)}_{13} \rangle - (i\Delta_{x}+\frac{\kappa_{x}}{2}) \langle \hat{a} \rangle,\\
\frac{d\langle \hat{\sigma}^{(k)}_{13} \rangle}{dt} = &~i g_{x} \langle \hat{a} \rangle (\langle \hat{\sigma}^{(k)}_{33} \rangle - \langle \hat{\sigma}^{(k)}_{11} \rangle) + \Omega_{y} \langle \hat{\sigma}^{(k)}_{23} \rangle - i \Omega_{z} \langle \hat{\sigma}^{(k)}_{12} \rangle 
\nonumber\\&
- (i \Delta_{13} + \Lambda + \frac{3\gamma_{l}}{2}+\Gamma) \langle \hat{\sigma}^{(k)}_{13} \rangle,\\
\frac{d\langle \hat{\sigma}^{(k)}_{12} \rangle}{dt} = &~\Omega_{y} (\langle \hat{\sigma}^{(k)}_{22} \rangle - \langle \hat{\sigma}^{(k)}_{11} \rangle) - i \Omega_{z}  \langle \hat{\sigma}^{(k)}_{13} \rangle + i g_{x} \langle \hat{a} \rangle \langle \hat{\sigma}^{(k)}_{32} \rangle
\nonumber\\&
- (i \Delta_{12} + \Lambda + \frac{3\gamma_{l}}{2}+\Gamma) \langle \hat{\sigma}^{(k)}_{12} \rangle,\\
\frac{d\langle \hat{\sigma}^{(k)}_{23} \rangle}{dt} = &~i \Omega_{z} (\langle \hat{\sigma}^{(k)}_{33} \rangle - \langle \hat{\sigma}^{(k)}_{22} \rangle)
- \Omega_{y} \langle \hat{\sigma}^{(k)}_{13} \rangle - i g_{x} \langle \hat{a} \rangle \langle \hat{\sigma}^{(k)}_{21} \rangle
\nonumber\\&
+\Gamma\langle \hat{\sigma}^{(k)}_{32} \rangle - (i\Delta_{23} + \gamma_{l} + \Lambda + \Gamma) \langle \hat{\sigma}^{(k)}_{23} \rangle,\\
\frac{d\langle \hat{\sigma}^{(k)}_{22} \rangle}{dt} = &~i \Omega_{z}  (\langle \hat{\sigma}^{(k)}_{32} \rangle - \langle \hat{\sigma}^{(k)}_{23} \rangle) - \Omega_{y} (\langle \hat{\sigma}^{(k)}_{12} \rangle + \langle \hat{\sigma}^{(k)}_{21} \rangle)
\nonumber\\&
+ \gamma_{l} \langle \hat{\sigma}^{(k)}_{11} \rangle - (\Lambda + \gamma_{l} + \Gamma) \langle \hat{\sigma}^{(k)}_{22} \rangle  + \Gamma \langle \hat{\sigma}^{(k)}_{33} \rangle
\nonumber\\&
+ \gamma_{25} \langle \hat{\sigma}^{(k)}_{55} \rangle + \gamma_{27} \langle \hat{\sigma}^{(k)}_{77} \rangle,\\
\frac{d\langle \hat{\sigma}^{(k)}_{33} \rangle}{dt} =&~i \Omega_{z} (\langle \hat{\sigma}^{(k)}_{23} \rangle - \langle \hat{\sigma}^{(k)}_{32} \rangle) + i g (\langle \hat{\sigma}^{(k)}_{13} \rangle \langle \hat{a}^{\dagger} \rangle - \langle \hat{a} \rangle \langle \hat{\sigma}^{(k)}_{31} \rangle) 
\nonumber\\&
+ \gamma_{l} \langle \hat{\sigma}^{(k)}_{11} \rangle + \Gamma \langle \hat{\sigma}^{(k)}_{22} \rangle- (\Lambda +\gamma_{l} + \Gamma) \langle \hat{\sigma}^{(k)}_{33} \rangle 
\nonumber\\&
+ \gamma_{36} \langle \hat{\sigma}^{(k)}_{66} \rangle + \gamma_{37} \langle \hat{\sigma}^{(k)}_{77} \rangle,\\
\frac{d\langle \hat{\sigma}^{(k)}_{44} \rangle}{dt} =&~\Lambda \langle \hat{\sigma}^{(k)}_{11} \rangle - (\gamma_{14} + \gamma_{74}) \langle \hat{\sigma}^{(k)}_{44} \rangle, \\
\frac{d\langle \hat{\sigma}^{(k)}_{55} \rangle}{dt} =&~\Lambda \langle \hat{\sigma}^{(k)}_{22} \rangle- (\gamma_{25} + \gamma_{75}) \langle \hat{\sigma}^{(k)}_{55} \rangle,\\
\frac{d\langle \hat{\sigma}^{(k)}_{66} \rangle}{dt} =&~\Lambda \langle \hat{\sigma}^{(k)}_{33} \rangle - (\gamma_{36} + \gamma_{76}) \langle \hat{\sigma}^{(k)}_{66} \rangle,\\ 
\frac{d\langle \hat{\sigma}^{(k)}_{77} \rangle}{dt} =&~\gamma_{74} \langle \hat{\sigma}^{(k)}_{44} \rangle + \gamma_{75} \langle \hat{\sigma}^{(k)}_{55} \rangle + \gamma_{76} \langle \hat{\sigma}^{(k)}_{66} \rangle
\nonumber\\&
- (\gamma_{17} + \gamma_{27} + \gamma_{37}) \langle \hat{\sigma}^{(k)}_{77} \rangle,
\end{align}
where $\Delta_{23}=\Delta_{13}-\Delta_{12}$. From equation (3) for the steady state condition,
\begin{align}
&G=~\frac{(g_{x}/\kappa_{x}n_{\mathrm{th}})}{\xi(1+\tilde{\Delta}^{2})}\sum_{k=1}^{N} \left(2g_{x} n_{x} (\langle \hat{\sigma}^{(k)}_{33} \rangle - \langle \hat{\sigma}^{(k)}_{11} \rangle)\right. 
\nonumber\\&
-\Omega_{y}((i+\tilde{\Delta})\langle \hat{a}^{\dagger}\rangle \langle \hat{\sigma}^{(k)}_{23} \rangle 
-(i-\tilde{\Delta}) \langle \hat{a} \rangle \langle \hat{\sigma}^{(k)}_{32} \rangle)
\nonumber\\&
- \Omega_{z} ((1-i\tilde{\Delta})\langle \hat{a}^{\dagger}\rangle \langle \hat{\sigma}^{(k)}_{12} \rangle
+ \left. (1+i\tilde{\Delta}) \langle \hat{a} \rangle \langle \hat{\sigma}^{(k)}_{21} \rangle)\right)
\end{align}
where $\xi=\Gamma+(3/2)\gamma_{l}+\Lambda$ and $\tilde{\Delta}= \Delta_{13}/\xi$.
The first line of equation (12) is the conventional gain term which is positive only if $\langle \hat{\sigma}^{(k)}_{33} \rangle > \langle \hat{\sigma}^{(k)}_{11} \rangle$, and the remaining two lines are the effects of the coherences $\langle \hat{\sigma}^{k}_{12} \rangle$ and $\langle \hat{\sigma}^{k}_{23} \rangle$ due to the MW and RF drive.

To verify the masing phase transition, we consider the second-order intensity correlation at zero delay given as $g^{(2)}(0) = \langle\hat{a}^\dagger\hat{a}^\dagger\hat{a}\hat{a}\rangle/|\langle\hat{a}^\dagger\hat{a}\rangle|^2$ 
At steady state, $g^{(2)}(0)$ can be expressed by reducing to first and second-order correlations as \cite{kubo1962generalized,plankensteiner2022quantumcumulants}\cite{Supplementary_information}, 
\begin{equation}
g^{(2)}(0) = \frac{|\langle \hat{a}\hat{a} \rangle|^2 + 2|\langle \hat{a}^\dagger\hat{a} \rangle|^2-2|\hat{a}|^4}{|\langle \hat{a}^\dagger\hat{a} \rangle|^2}.
\label{eq:g2}
\end{equation}

The terms with phase, such as $\langle \hat{a}\rangle$ and $\langle \hat{a}\hat{a} \rangle$, vanish in a conventional laser due to the phase-invariant condition, and the expression in equation (\ref{eq:g2}) only gives a value of 2.
The phase-invariant condition does not hold in our model, as the phase of the photon operator is influenced by the additional MW and RF drives, and we use equation (\ref{eq:g2}).
The phase invariant condition is further confirmed in our numerical simulations presented below, as even if we consider $\langle\hat{a}\rangle=0$ as an initial condition to solve the transient coupled equations in Eqs. (2)–(11), the resultant $n_{x}=\langle\hat{a}^{\dagger}\rangle \langle\hat{a}\rangle$ is non-zero (refer to the code and transient solution in \cite{Supplementary_information}).

\paragraph*{Numerical estimation \textemdash} We numerically model the system to demonstrate the feasibility of the model.
For this, we keep only $\Lambda$, $\omega_{y}$, $\Omega_{y}$, $\omega_{z}$, and $\Omega_{z}$ as free parameters within experimentally achievable ranges and fix all the other parameters.
We consider experimentally relevant values for the fixed parameters $g/2\pi = 18~\text{mHz}$, $N = 1.1\times 10^{14}$, $\kappa_x/2\pi = 130~\text{kHz}$, $\Gamma/2\pi = 330~\text{kHz}$~\cite{wang2024spin} (Note that the temperature in \cite{wang2024spin} is 293~K; here we assume 300~K for simplicity), and $\gamma_{l}/2\pi = 200~\text{Hz}$~\cite{jarmola2012temperature}.
All the numerical values used for fixed parameters in our system, including, intrinsic NV rates, are tabulated in \cite{Supplementary_information}.

We consider $\Lambda/2\pi$ between 0 to 1000~Hz.
For a laser spot-size of $A=0.09$~cm$^2$ and absorption cross-section of $\sigma=7.8\times10^{-17}$~cm$^2$ as in reference \cite{wang2024spin}, the optical power $P=\Lambda \hbar \nu A/\sigma$, where $\nu$ is the optical excitation frequency, $\Lambda/2\pi = 1000$~Hz, corresponds to laser powers up to 0.4~W, which is experimentally achievable.
For both $\Omega_{y}/2\pi$ and $\Omega_{z}/2\pi$, we consider the numerical values between 0 to 10~MHz.
This range for homogeneous classical MW driving of the NV spin ensemble can be achieved with special MW antennas \cite{bayat2014efficient, yaroshenko2020circularly, opaluch2021optimized, ben2024modified, rezinkin2024uniform}.
On the other hand, $\Omega_{z}/2\pi \sim ~ 1$~MHz has been experimentally shown for an ensemble of NV spins in reference \cite{mikawa2023electron}.

To estimate $n_{x}$, we numerically solve the coupled equations in Eqs. (2)–(11) in steady state conditions using a standard iterative Newton–Raphson root-finding solver, assuming $\langle \hat{\sigma}^{(i)}_{\alpha\beta} \rangle = \langle \hat{\sigma}^{(k)}_{\alpha\beta} \rangle$, and replacing the summation in the equations with $N$.
For estimating $g^{(2)}(0)$ and checking $n_x$ obtained from equations Eqs. (2)–(11), we also consider the second-order case for our system using \texttt{QuantumCumulants.jl}, \texttt{ModellingToolkit.jl}, and \texttt{DifferentialEquations.jl} \cite{Supplementary_information}.
We numerically solve a closed set of 75 equations in total for the second-order case in Julia up to a time of 0.5~ms for the system to reach a steady state using the Bogacki-Shampine 3 (BS3) Runge-Kutta method.

\begin{figure}
    \centering
    \includegraphics[width=\linewidth]{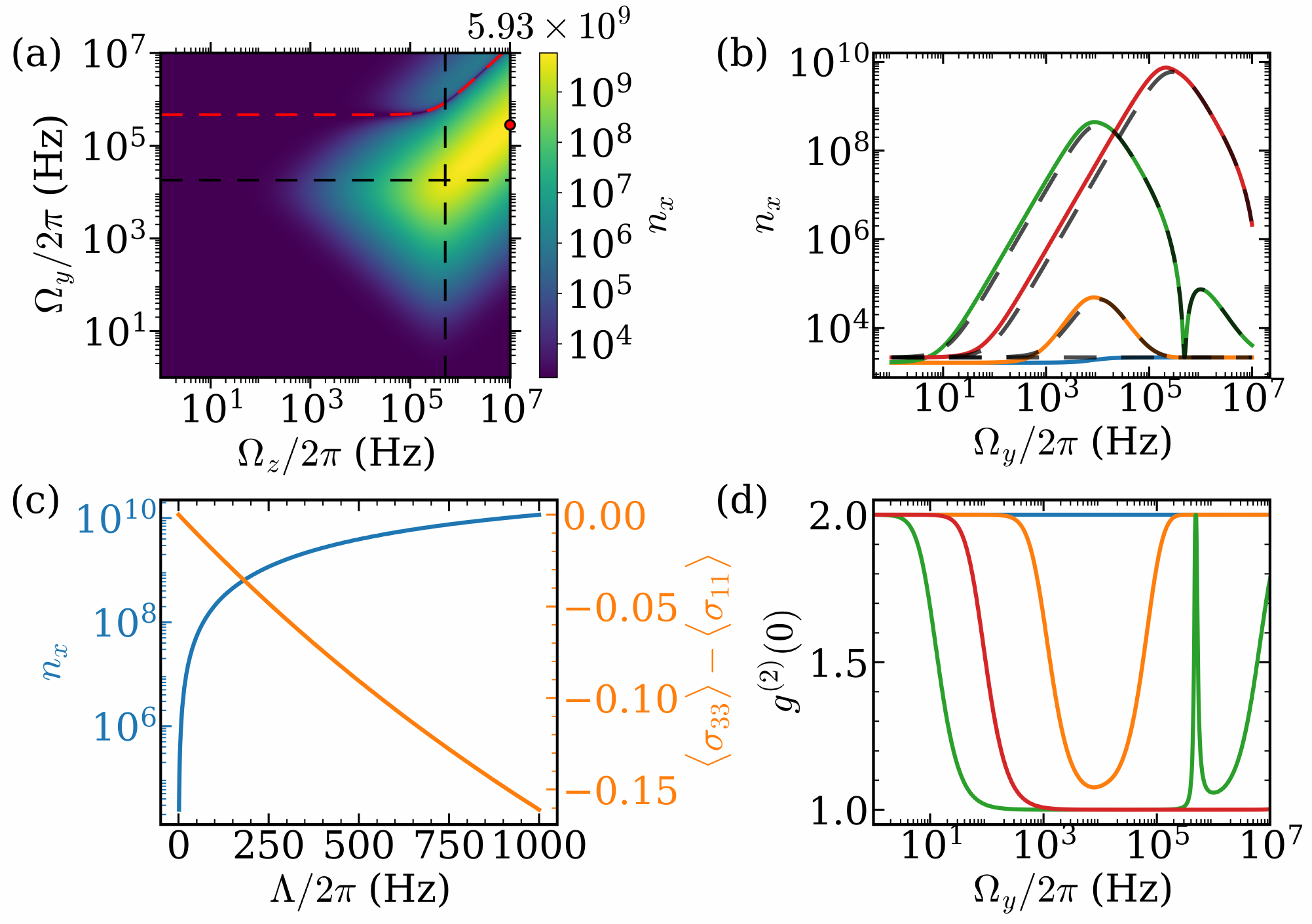}
    \caption{Numerical results. (a) Steady-state photon number $n_{x}$ as a function of $\Omega_{y}$ and $\Omega_{z}$ for $\Lambda/2\pi=500$~Hz. The maximum $n_x$ occurs at $\Omega_z/2\pi=10$~MHz and $\Omega_y/2\pi=282$~kHz, indicated by the red dot. The black horizontal and vertical dashed lines indicate the driving strengths at $\Omega_y/2\pi=18$~kHz and $\Omega_z/2\pi=500$~kHz, respectively. A minimum $n_x$ region is observed along the dashed red line, where the gain effects due to the double driving scheme are minimum (see \cite{Supplementary_information} for more details)
    (b) $n_x$ using second-order correlation equations as a function of $\Omega_y$ for different $\Omega_z/2\pi=10$~Hz (blue), 1~kHz (orange), 100~kHz (green), 10~MHz (red), showing agreement with the first-order equations from (a) (black-dashed lines). 
    For the first-order equations, the thermal photon occupation number is added to the calculated photon number, since it is not included explicitly in the model.
    (c) Photon number $n_x$ (blue) and spin polarization (orange) as a function of $\Lambda$ at $\Omega_y/2\pi=18$~kHz and $\Omega_z/2\pi=500$~kHz.
    (d)  Second-order coherence function $g^{(2)}(0)$ plotted as a function of the driving strength $\Omega_{y}$ for different values of $\Omega_{z}/2\pi=10$~Hz (blue), 10$^3$~Hz (orange), 10$^5$~Hz (green), 10$^7$~Hz (red).}
    \label{fig:2}
\end{figure}

We present the numerical results in Figure~\ref{fig:2} for $\Delta_{x}=\Delta_{12}=\Delta_{13}=0$.
In Figure~\ref{fig:2}(a), we show $n_{x}$ as a function of $\Omega_{y}$ and $\Omega_{z}$ for a representative case of $\Lambda/2\pi = 500\text{Hz}$.
We can see that near the vertical (horizontal) dashed line, which corresponds to around $\Omega_{z}/2\pi = 500~\text{kHz}$ ($\Omega_{y}/2\pi = 18~\text{kHz}$), the large photon number region can be reached with minimal $\Omega_{y}/2\pi$($\Omega_{z}/2\pi$) at which $n_x\approx3.7\times10^9$ photons, giving a maser output power of approximately $-60$~dBm ($1$~nW).
In Figure~\ref{fig:2}(b), we compare and verify $n_{x}$ values obtained using first-order equations with values obtained using second-order equations.
We attribute the agreement between both sets of results to the fact that conventional cooperativity $C=4g_{x}^{2}N/(\kappa_{x}\Gamma) \sim 3$, which is not too large, and the first-order equations are sufficient to explain the key effects considered in this work within our approximations.
From Figure~\ref{fig:2}(c), we can see that by optical pumping, $n_{x}$ can be increased for realistic values of $\Omega_{y}$ and $\Omega_{z}$, though the optical pumping increases the population in $|1\rangle$.
We estimate $g^{(2)}(0)$ as shown in Figure~\ref{fig:2}(d) by solving second-order equations as mentioned above and see a phase transition from the thermal state ({$g^{(2)}(0)=2$}) to a coherent state ($g^{(2)}(0)=1$) in the region where the maximum $n_{x}$ occurs in Fig.\ref{fig:2}(a).

\paragraph*{Magnetic-field sensing \textemdash}The inversionless maser can be used to detect a weak magnetic field along the $x$-axis, through the change in the maser output intensity. To find the optimal sensing point, we tune the $B_x$ field with an additional $x$-field denoted by $B_\mathrm{s}$. $B_{\mathrm{s}}$ modifies $n_{x}$ through $\Delta_{12}$ and $\Delta_{23}$ with $\Delta_{23} = \Delta_{13}-\Delta_{12}$ and the basis considered in this work remains the same for $\gamma_e(B_x+B_\mathrm{s})\ll D$ and $B_\mathrm{s}\ll B_x$, where $B_x=5$~mT.
Assuming that the maser output is detected in the same way as in reference \cite{eisenach2021cavity}, the best possible Johnson-Nyquist noise limited magnetic field sensitivity ($\eta_B$) can be expressed as, $\eta_B = \sqrt{k_B T n_x}/(\sqrt{\hbar\omega_x\kappa_x}|R_0|)$ (see supplementary information of \cite{eisenach2021cavity}).
Here $R_0=d n_{x} / d \mathrm{B}_{\mathrm{s}}$ is the slope of the curve $n_{x}$ as a function of $B_{\mathrm{s}}$, and $T$ is the device temperature. 
In an actual experimental scenario, the device temperature is expected to be greater than 300~K due to the optical pump, MW drive, and RF drive.
For example, in \cite{wang2024spin}, the sensitivity is estimated for the noise floor at $T=407$~K.
We do not focus on determining an absolute value of $\eta_B$, rather we are interested in the order of magnitude and comparing it with the standard quantum limit (SQL), $\eta_{\text{SQL}}=1/(\gamma_{e}(N T_{2}^{*})^{1/2})$, which for our case is $\sim$ fT/$\sqrt{\mathrm{Hz}}$ \cite{wang2024spin}.

\begin{figure}
    \centering
    \includegraphics[width=\linewidth]{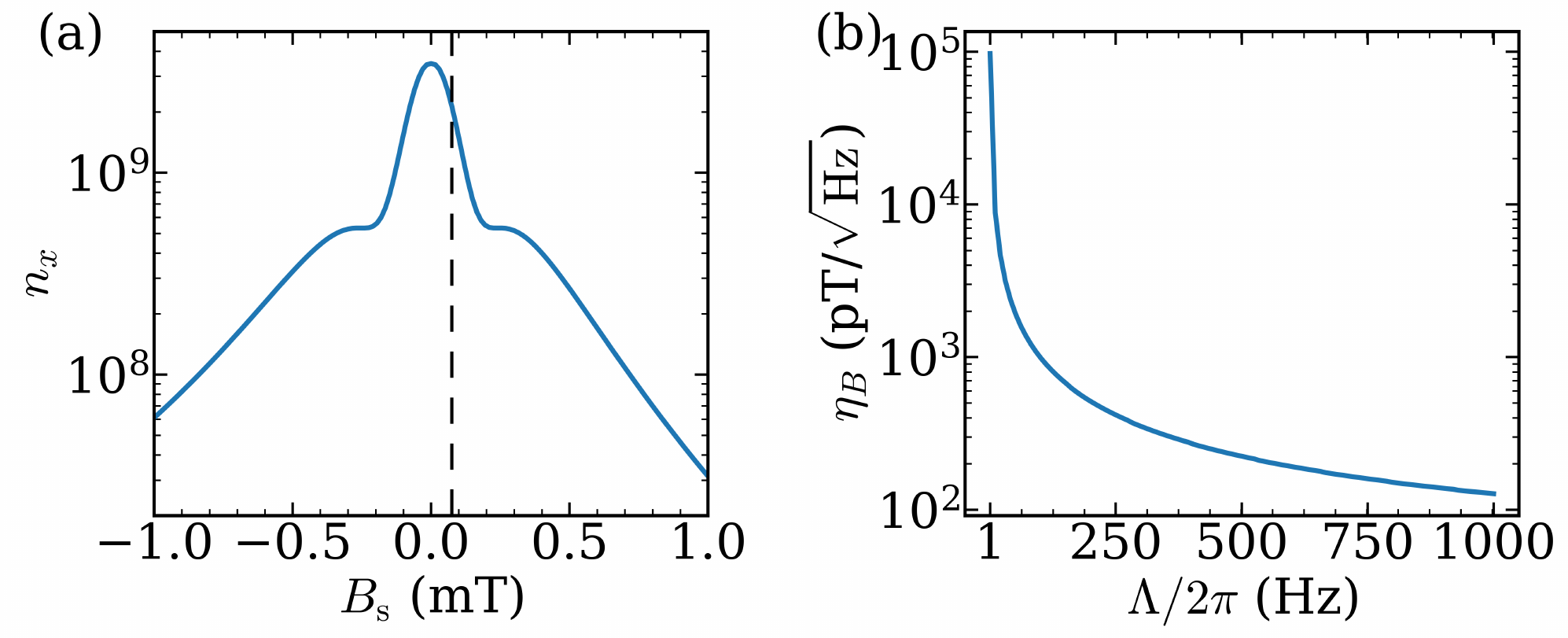}
    \caption{Magnetic field sensing. (a) Photon number $n_x$ as a function of additional $B_x$ field denoted by $B_\mathrm{s}$, evaluated at driving field strengths $\Omega_y/2\pi = 33.5$~kHz, $\Omega_z/2\pi=628$~kHz, pump rate $\Lambda/2\pi = 500$~Hz (optimal driving strength for sensitivity, refer to \cite{Supplementary_information} for more details). The maximum slope $|R_0|$ occurs near $B_{\mathrm{s}} = 0.06~\mathrm{mT}$, indicated by the vertical dashed line. (b) Magnetic sensitivity as a function of pump rate at the operating point in (a).}
    \label{fig:3}
\end{figure}
To show the sensing capability, we plot $n_{x}$, as a function of $B_{s}$ in Figure \ref{fig:3} (a), where $\Delta_{x}=\Delta_{12}=\Delta_{23}=0$.
Here we chose $\Omega_y/2\pi = 33.5$~kHz, $\Omega_z/2\pi=628$~kHz as we find these values are expected to give the best sensitivity \cite{Supplementary_information}.
The optimum sensitivity is obtained for the $B_{\mathrm{s}}$ corresponding to the dashed vertical line in Figure \ref{fig:3} (a), and we plot this optimum sensitivity as a function of $\Lambda/2\pi$ in Figure \ref{fig:3} (b).
We obtain a sensitivity of the order of 100~pT/$\sqrt{\mathrm{Hz}}$, which is almost two orders of magnitude higher than the one in reference \cite{wang2024spin} and five orders of magnitude higher than the SQL.
Although the sensitivity can be slightly improved by tuning $\Delta_{12}$ and $\Delta_{23}$ for $B_{\mathrm{s}} = 0$, it does not lead to an order-of-magnitude improvement \cite{Supplementary_information}. 
We also verify the sensitivity with second-order equations and obtain results similar to the first-order equations, as expected \cite{Supplementary_information}.

\paragraph*{Conclusion \textemdash}In conclusion, we propose a method supported by experimentally feasible numerical estimates for realizing a room-temperature, $\sim2.9$~GHz, diamond NV electronic spin maser without the need for population inversion and a strong bias magnetic field, as required in conventional NV masers.
We also show that the proposed inversionless maser can be used as a magnetic-field sensor by measuring changes in the resonant maser output intensity as a function of the external magnetic field. Although the sensitivity is approximately five orders of magnitude above the SQL, it opens up the possibility of a diamond maser magnetometer with a lighter and smaller sensor head.
Our results could have impactful applications in the development of scalable quantum devices.

\begin{acknowledgments}
A.F. acknowledges funding through a domestic Macquarie University (MQ) and Sydney Quantum Academy (SQA) scholarship.
\end{acknowledgments}

\vspace{-10pt}

\section*{Authors' contributions}
SR developed the idea of the project and the methodology. AF derived the equations with contributions from SR. AF performed the numerical modeling and analysis with feedback from SR. SR and AF wrote the manuscript. SR supervised the project.

\bibliography{references.bib}

@misc{Supplementary_information,
    note={Supplementary information for `Room-temperature inversionless diamond nitrogen-vacancy electronic spin maser'}
}

@article{jin2015proposal,
  title={Proposal for a room-temperature diamond maser},
  author={Jin, Liang and Pfender, Matthias and Aslam, Nabeel and Neumann, Philipp and Yang, Sen and Wrachtrup, J{\"o}rg and Liu, Ren-Bao},
  journal={Nature Communications},
  volume={6},
  number={1},
  pages={8251},
  year={2015},
  publisher={Nature Publishing Group UK London},
  doi={https://doi.org/10.1038/ncomms9251}
}

@article{breeze2018continuous,
  title={Continuous-wave room-temperature diamond maser},
  author={Breeze, Jonathan D and Salvadori, Enrico and Sathian, Juna and Alford, Neil McN and Kay, Christopher WM},
  journal={Nature},
  volume={555},
  number={7697},
  pages={493--496},
  year={2018},
  publisher={Nature Publishing Group UK London},
  doi={https://doi.org/10.1038/nature25970}  
}

@article{zollitsch2023maser,
  title={Maser threshold characterization by resonator Q-factor tuning},
  author={Zollitsch, Christoph W and Ruloff, Stefan and Fett, Yan and Wiedemann, Haakon TA and Richter, Rudolf and Breeze, Jonathan D and Kay, Christopher WM},
  journal={Communications Physics},
  volume={6},
  number={1},
  pages={295},
  year={2023},
  publisher={Nature Publishing Group UK London},
  doi={https://doi.org/10.1038/s42005-023-01418-3}
}

@article{sherman2022diamond,
  title={Diamond-based microwave quantum amplifier},
  author={Sherman, Alexander and Zgadzai, Oleg and Koren, Boaz and Peretz, Idan and Laster, Eyal and Blank, Aharon},
  journal={Science Advances},
  volume={8},
  number={49},
  pages={eade6527},
  year={2022},
  publisher={American Association for the Advancement of Science},
  doi={DOI: 10.1126/sciadv.ade6527}
}

@article{day2024room,
  title={Room-temperature solid-state maser amplifier},
  author={Day, Tom and Isarov, Maya and Pappas, William J and Johnson, Brett C and Abe, Hiroshi and Ohshima, Takeshi and McCamey, Dane R and Laucht, Arne and Pla, Jarryd J},
  journal={Physical Review X},
  volume={14},
  number={4},
  pages={041066},
  year={2024},
  publisher={APS},
  doi={https://doi.org/10.1103/PhysRevX.14.041066}
}

@article{ng2025portable,
  title={Portable maser oscillator at room temperature with reduced magnetic field requirements through spatial orientation},
  author={Ng, Wern and Wen, Yongqiang and Alford, Neil McN and Arroo, Daan M},
  journal={Physical Review Applied},
  volume={23},
  number={5},
  pages={054064},
  year={2025},
  publisher={APS},
  doi={https://doi.org/10.1103/PhysRevApplied.23.054064}
}

@article{arroo2021perspective,
  title={Perspective on room-temperature solid-state masers},
  author={Arroo, Daan M and Alford, Neil McN and Breeze, Jonathan D},
  journal={Applied Physics Letters},
  volume={119},
  number={14},
  year={2021},
  publisher={AIP Publishing},
  doi={https://doi.org/10.1063/5.0061330}
}

@article{wu2022enhanced,
  title={Enhanced quantum sensing with room-temperature solid-state masers},
  author={Wu, Hao and Yang, Shuo and Oxborrow, Mark and Jiang, Min and Zhao, Qing and Budker, Dmitry and Zhang, Bo and Du, Jiangfeng},
  journal={Science advances},
  volume={8},
  number={48},
  pages={eade1613},
  year={2022},
  publisher={American Association for the Advancement of Science},
  doi={DOI: 10.1126/sciadv.ade1613}
}

@article{jeske2016laser,
  title={Laser threshold magnetometry},
  author={Jeske, Jan and Cole, Jared H and Greentree, Andrew D},
  journal={New Journal of Physics},
  volume={18},
  number={1},
  pages={013015},
  year={2016},
  publisher={IOP Publishing},
  doi={10.1088/1367-2630/18/1/013015}
}

@article{mompart2000lasing,
  title={Lasing without inversion},
  author={Mompart, J and Corbalan, R},
  journal={Journal of Optics B: Quantum and Semiclassical Optics},
  volume={2},
  number={3},
  pages={R7--R24},
  year={2000},
  doi={10.1088/1464-4266/2/3/201}
}

@article{wang2025cavity,
  title={Cavity-enhanced solid-state nuclear spin gyroscope},
  author={Wang, Hanfeng and Wu, Shuang and Jacobs, Kurt and Duan, Yuqin and Englund, Dirk R and Trusheim, Matthew E},
  journal={Physical Review Letters},
  volume={134},
  number={18},
  pages={183603},
  year={2025},
  publisher={APS},
  doi={https://doi.org/10.1103/PhysRevLett.134.183603}
}

@article{tetienne2012magnetic,
  title={Magnetic-field-dependent photodynamics of single NV defects in diamond: an application to qualitative all-optical magnetic imaging},
  author={Tetienne, Jean Philippe and Rondin, Lo{\"\i}c and Spinicelli, Piernicola and Chipaux, Mayeul and Debuisschert, Thierry and Roch, Jean-Fran{\c{c}}ois and Jacques, Vincent},
  journal={New Journal of Physics},
  volume={14},
  number={10},
  pages={103033},
  year={2012},
  publisher={IOP Publishing},
  doi={10.1088/1367-2630/14/10/103033}
}

@article{lamba2024vector,
  title={Vector detection of ac magnetic fields by nitrogen vacancy centers of single orientation in diamond},
  author={Lamba, Pooja and Rana, Akshat and Halder, Sougata and Dhomkar, Siddharth and Suter, Dieter and Kamineni, Rama K},
  journal={Physical Review B},
  volume={109},
  number={19},
  pages={195424},
  year={2024},
  publisher={APS},
  doi={https://doi.org/10.1103/PhysRevB.109.195424}
}

@article{gruber1997scanning,
  title={Scanning confocal optical microscopy and magnetic resonance on single defect centers},
  author={Gruber, A and Drabenstedt, A and Tietz, C and Fleury, L and Borczyskowski, C von},
  journal={Science},
  volume={276},
  number={5321},
  pages={2012--2014},
  year={1997},
  publisher={American Association for the Advancement of Science},
  doi={https://doi.org/10.1126/science.276.5321.2012}
}

@article{doherty2011negatively,
  title={The negatively charged nitrogen-vacancy centre in diamond: the electronic solution},
  author={Doherty, Marcus W and Manson, Neil B and Delaney, Paul and Hollenberg, Lloyd CL},
  journal={New Journal of Physics},
  volume={13},
  number={2},
  pages={025019},
  year={2011},
  doi = {10.1088/1367-2630/13/2/025019}
}

@article{doherty2013nitrogen,
  title={The nitrogen-vacancy colour centre in diamond},
  author={Doherty, Marcus W and Manson, Neil B and Delaney, Paul and Jelezko, Fedor and Wrachtrup, J{\"o}rg and Hollenberg, Lloyd CL},
  journal={Physics Reports},
  volume={528},
  number={1},
  pages={1--45},
  year={2013},
  publisher={Elsevier},
  doi={https://doi.org/10.1016/j.physrep.2013.02.001}
}

@article{gupta2016efficient,
  title={Efficient signal processing for time-resolved fluorescence detection of nitrogen-vacancy spins in diamond},
  author={Gupta, Anchal and Hacquebard, Luke and Childress, Lilian},
  journal={Journal of the Optical Society of America B},
  volume={33},
  number={3},
  pages={B28--B34},
  year={2016},
  publisher={Optical Society of America},
  doi={https://doi.org/10.1364/JOSAB.33.000B28}
}

@article{rogers2008infrared,
  title={Infrared emission of the NV centre in diamond: Zeeman and uniaxial stress studies},
  author={Rogers, LJ and Armstrong, S and Sellars, MJ and Manson, NB},
  journal={New Journal of Physics},
  volume={10},
  number={10},
  pages={103024},
  year={2008},
  doi={https://doi.org/10.1088/1367-2630/10/10/103024}
}

@article{acosta2010optical,
  title={Optical properties of the nitrogen-vacancy singlet levels in diamond},
  author={Acosta, VM and Jarmola, A and Bauch, E and Budker, D},
  journal={Physical Review B—Condensed Matter and Materials Physics},
  volume={82},
  number={20},
  pages={201202},
  year={2010},
  publisher={APS},
  doi={https://doi.org/10.1103/PhysRevB.82.201202}
}

@article{zhang2022cavity,
  title={Cavity quantum electrodynamics effects with nitrogen vacancy center spins coupled to room temperature microwave resonators},
  author={Zhang, Yuan and Wu, Qilong and Su, Shi-Lei and Lou, Qing and Shan, Chongxin and M{\o}lmer, Klaus},
  journal={Physical review letters},
  volume={128},
  number={25},
  pages={253601},
  year={2022},
  publisher={APS},
  doi={https://doi.org/10.1103/PhysRevLett.128.253601}
}

@article{zhang2022microwave,
  title={Microwave mode cooling and cavity quantum electrodynamics effects at room temperature with optically cooled nitrogen-vacancy center spins},
  author={Zhang, Yuan and Wu, Qilong and Wu, Hao and Yang, Xun and Su, Shi-Lei and Shan, Chongxin and M{\o}lmer, Klaus},
  journal={npj Quantum Information},
  volume={8},
  number={1},
  pages={125},
  year={2022},
  publisher={Nature Publishing Group UK London},
  doi={https://doi.org/10.1038/s41534-022-00642-z}
}

@article{plankensteiner2022quantumcumulants,
  title={QuantumCumulants. jl: A Julia framework for generalized mean-field equations in open quantum systems},
  author={Plankensteiner, David and Hotter, Christoph and Ritsch, Helmut},
  journal={Quantum},
  volume={6},
  pages={617},
  year={2022},
  publisher={Verein zur F{\"o}rderung des Open Access Publizierens in den Quantenwissenschaften},
  doi={https://doi.org/10.22331/q-2022-01-04-617}
}

@article{wang2024spin,
  title={A spin-refrigerated cavity quantum electrodynamic sensor},
  author={Wang, Hanfeng and Tiwari, Kunal L and Jacobs, Kurt and Judy, Michael and Zhang, Xin and Englund, Dirk R and Trusheim, Matthew E},
  journal={Nature Communications},
  volume={15},
  number={1},
  pages={10320},
  year={2024},
  publisher={Nature Publishing Group UK London},
  doi={https://doi.org/10.1038/s41467-024-54333-8}
}

@article{jarmola2012temperature,
  title={Temperature-and magnetic-field-dependent longitudinal spin relaxation in nitrogen-vacancy ensembles in diamond},
  author={Jarmola, A and Acosta, VM and Jensen, K and Chemerisov, S and Budker, D},
  journal={Physical review letters},
  volume={108},
  number={19},
  pages={197601},
  year={2012},
  publisher={APS},
  doi={https://doi.org/10.1103/PhysRevLett.108.197601}
}

@article{eisenach2021cavity,
  title={Cavity-enhanced microwave readout of a solid-state spin sensor},
  author={Eisenach, Erik R and Barry, John F and O’Keeffe, Michael F and Schloss, Jennifer M and Steinecker, Matthew H and Englund, Dirk R and Braje, Danielle A},
  journal={Nature communications},
  volume={12},
  number={1},
  pages={1357},
  year={2021},
  publisher={Nature Publishing Group UK London},
  doi={https://doi.org/10.1038/s41467-021-21256-7}
}

@article{bayat2014efficient,
  title={Efficient, uniform, and large area microwave magnetic coupling to NV centers in diamond using double split-ring resonators},
  author={Bayat, Khadijeh and Choy, Jennifer and Farrokh Baroughi, Mahdi and Meesala, Srujan and Loncar, Marko},
  journal={Nano letters},
  volume={14},
  number={3},
  pages={1208--1213},
  year={2014},
  publisher={ACS Publications},
  doi={https:/doi.org/10.1021/nl404072s}
}

@article{yaroshenko2020circularly,
  title={Circularly polarized microwave antenna for nitrogen vacancy centers in diamond},
  author={Yaroshenko, Vitaly and Soshenko, Vladimir and Vorobyov, Vadim and Bolshedvorskii, Stepan and Nenasheva, Elizaveta and Kotel’nikov, Igor and Akimov, Alexey and Kapitanova, Polina},
  journal={Review of Scientific Instruments},
  volume={91},
  number={3},
  year={2020},
  publisher={AIP Publishing},
  doi={https://doi.org/10.1063/1.5129863}
}

@article{opaluch2021optimized,
  title={Optimized planar microwave antenna for nitrogen vacancy center based sensing applications},
  author={Opaluch, Oliver Roman and Oshnik, Nimba and Nelz, Richard and Neu, Elke},
  journal={Nanomaterials},
  volume={11},
  number={8},
  pages={2108},
  year={2021},
  publisher={MDPI},
  doi={https://doi.org/10.3390/nano11082108}
}

@article{ben2024modified,
  title={Modified Split-Ring Resonators for Efficient and Homogeneous Microwave Control of Large Volume Spin Ensembles},
  author={Ben-Shalom, Yachel and Hen, Amir and Bar-Gill, Nir},
  journal={IEEE Sensors Journal},
  volume={24},
  number={13},
  pages={20420--20426},
  year={2024},
  publisher={IEEE},
  doi={https://doi.org/10.1109/JSEN.2024.3401049}
}

@article{rezinkin2024uniform,
  title={Uniform microwave field formation for control of ensembles of negatively charged nitrogen vacancy in diamond},
  author={Rezinkin, Oleg and Rezinkina, Marina and Kitamura, Takuya and Paul, Rajan and Jelezko, Fedor},
  journal={Review of Scientific Instruments},
  volume={95},
  number={10},
  year={2024},
  publisher={AIP Publishing},
  doi={https://doi.org/10.1063/5.0203162}
}

@article{matsuzaki2016optically,
  title={Optically detected magnetic resonance of high-density ensemble of NV- centers in diamond},
  author={Matsuzaki, Yuichiro and Morishita, Hiroki and Shimooka, Takaaki and Tashima, Toshiyuki and Kakuyanagi, Kosuke and Semba, Kouichi and Munro, WJ and Yamaguchi, Hiroshi and Mizuochi, Norikazu and Saito, Shiro},
  journal={Journal of Physics: Condensed Matter},
  volume={28},
  number={27},
  pages={275302},
  year={2016},
  publisher={IOP Publishing},
  doi={https://doi.org/10.1088/0953-8984/28/27/275302}
}

@article{saijo2018ac,
  title={AC magnetic field sensing using continuous-wave optically detected magnetic resonance of nitrogen-vacancy centers in diamond},
  author={Saijo, Soya and Matsuzaki, Yuichiro and Saito, Shiro and Yamaguchi, Tatsuma and Hanano, Ikuya and Watanabe, Hideyuki and Mizuochi, Norikazu and Ishi-Hayase, Junko},
  journal={Applied Physics Letters},
  volume={113},
  number={8},
  year={2018},
  publisher={AIP Publishing},
  doi={https://doi.org/10.1063/1.5024401}
}

@article{yamaguchi2019bandwidth,
  title={Bandwidth analysis of AC magnetic field sensing based on electronic spin double-resonance of nitrogen-vacancy centers in diamond},
  author={Yamaguchi, Tatsuma and Matsuzaki, Yuichiro and Saito, Shiro and Saijo, Soya and Watanabe, Hideyuki and Mizuochi, Norikazu and Ishi-Hayase, Junko},
  journal={Japanese journal of applied physics},
  volume={58},
  number={10},
  pages={100901},
  year={2019},
  publisher={IOP Publishing},
  doi={https://doi.org/10.7567/1347-4065/ab3d03}
}

@article{tabuchi2023temperature,
  title={Temperature sensing with RF-dressed states of nitrogen-vacancy centers in diamond},
  author={Tabuchi, Hibiki and Matsuzaki, Yuichiro and Furuya, Noboru and Nakano, Yuta and Watanabe, Hideyuki and Tokuda, Norio and Mizuochi, Norikazu and Ishi-Hayase, Junko},
  journal={Journal of Applied Physics},
  volume={133},
  number={2},
  year={2023},
  publisher={AIP Publishing},
  doi={https://doi.org/10.1063/5.0129706}
}

@article{mikawa2023electron,
  title={Electron-spin double resonance of nitrogen-vacancy centers in diamond under a strong driving field},
  author={Mikawa, Takumi and Okaniwa, Ryusei and Matsuzaki, Yuichiro and Tokuda, Norio and Ishi-Hayase, Junko},
  journal={Physical Review A},
  volume={108},
  number={1},
  pages={012610},
  year={2023},
  publisher={APS},
  doi={https://doi.org/10.1103/PhysRevA.108.012610
Export Citation}
}

@article{shin2013suppression,
  title={Suppression of electron spin decoherence of the diamond NV center by a transverse magnetic field},
  author={Shin, Chang S and Avalos, Claudia E and Butler, Mark C and Wang, Hai-Jing and Seltzer, Scott J and Liu, Ren-Bao and Pines, Alexander and Bajaj, Vikram S},
  journal={Physical Review B—Condensed Matter and Materials Physics},
  volume={88},
  number={16},
  pages={161412},
  year={2013},
  publisher={APS},
  doi={https://doi.org/10.1103/PhysRevB.88.161412
Export Citation}
}

@article{kubo1962generalized,
  title={Generalized cumulant expansion method},
  author={Kubo, Ryogo},
  journal={Journal of the Physical Society of Japan},
  volume={17},
  number={7},
  pages={1100--1120},
  year={1962},
  publisher={The Physical Society of Japan}
}

@article{harris1989lasers,
  title={Lasers without inversion: Interference of lifetime-broadened resonances},
  author={Harris, Stephen E},
  journal={Physical review letters},
  volume={62},
  number={9},
  pages={1033},
  year={1989},
  publisher={APS}
}

@article{RevModPhys.84.621,
  title = {Gaussian quantum information},
  author = {Weedbrook, Christian and Pirandola, Stefano and Garc\'{\i}a-Patr\'on, Ra\'ul and Cerf, Nicolas J. and Ralph, Timothy C. and Shapiro, Jeffrey H. and Lloyd, Seth},
  journal = {Rev. Mod. Phys.},
  volume = {84},
  issue = {2},
  pages = {621--669},
  numpages = {0},
  year = {2012},
  month = {May},
  publisher = {American Physical Society},
  doi = {10.1103/RevModPhys.84.621},
  url = {https://link.aps.org/doi/10.1103/RevModPhys.84.621}
}

@article{werren2022quantum,
  title={A quantum model of lasing without inversion},
  author={Werren, Nicholas and Gauger, Erik M and Kirton, Peter},
  journal={New Journal of Physics},
  volume={24},
  number={9},
  pages={093027},
  year={2022},
  publisher={IOP Publishing},
  doi = {10.1088/1367-2630/ac8d27}
}

@article{saldutti2024onset,
  title={The onset of lasing in semiconductor nanolasers},
  author={Saldutti, Marco and Yu, Yi and M{\o}rk, Jesper},
  journal={Laser \& Photonics Reviews},
  volume={18},
  number={11},
  pages={2300840},
  year={2024},
  publisher={Wiley Online Library},
  doi = {10.1002/lpor.202300840}
}

@article{PhysRevA.50.4318,
  title = {Photon statistics of a cavity-QED laser: A comment on the laser--phase-transition analogy},
  author = {Rice, Perry R. and Carmichael, H. J.},
  journal = {Phys. Rev. A},
  volume = {50},
  issue = {5},
  pages = {4318--4329},
  numpages = {0},
  year = {1994},
  month = {Nov},
  publisher = {American Physical Society},
  doi = {10.1103/PhysRevA.50.4318},
  url = {https://link.aps.org/doi/10.1103/PhysRevA.50.4318}
}

\end{document}


\title{Supplementary information for `Room-temperature inversionless diamond nitrogen-vacancy electronic spin maser'}

\author{\orcidlink{0009-0007-7498-0741}Ali Fawaz}
\affiliation{School of Mathematical and Physical Sciences, Macquarie University, North Ryde, NSW 2109, Australia}

\author{\orcidlink{0000-0002-9081-5750}Sarath Raman Nair}
\email[Corresponding author]{}

\affiliation{School of Mathematical and Physical Sciences, Macquarie University, North Ryde, NSW 2109, Australia}

\maketitle

\date{\today}

\section{Derivation of Rotating Frame Hamiltonian}
\setcounter{equation}{0}
\setcounter{figure}{0}
Starting with the ground state spin Hamiltonian for a single NV spin (spin-1) interacting with an external magnetic field $\mathbf{B} = (B_x,B_y,B_z)$, and neglecting any strain-induced splitting and the hyperfine coupling to nuclear spins as \cite{tetienne2012magnetic},
\begin{equation}
    H = \hbar D \mathbf{\hat{S}}_z^2+\hbar\gamma_{e}\mathbf{B}\cdot \mathbf{S}
\end{equation}
where $\mathbf{S}=(\mathbf{\hat{S}}_x,\mathbf{\hat{S}}_y,\mathbf{\hat{S}}_z$) and $\mathbf{\hat{S}}_i$ for $i\in {x,y,z}$ are the spin-1 operators. $\gamma_e$ is the electron gyromagnetic ratio, and $D$ is the zero-field splitting of the NV spin.

With the static and driving fields described in the main text, the Hamiltonian can be expressed as follows:

\begin{equation}
    H = \hbar D \mathbf{\hat{S}}_z^2+\hbar\gamma_{e}B_x \mathbf{\hat{S}_x} + \hbar\gamma_{e}B_{y}\cos(\omega_{y}t) \mathbf{\hat{S}_y} + \hbar\gamma_{e}B_{z}\cos(\omega_{z}t) \mathbf{\hat{S}_z} 
    \label{eq:S.I.2}
\end{equation}

The Hamiltonian above includes interactions with three magnetic fields (one static, two AC fields). The static field is along the $x$-direction ($B_x$), the microwave (MW) field is along the y-direction ($B_{y}\cos(\omega_{y}t)$), and the radio-frequency (RF) field is along the $z$-direction ($B_{z}\cos(\omega_{z}t)$).

To find an appropriate basis, we focus on the first two terms of the Hamiltonian in Eq. (\ref{eq:S.I.2}). Diagonalizing this part of the Hamiltonian yields:

\begin{equation}
    \hat{H}_{NV} = \hbar D \mathbf{\hat{S}}_z^2+\hbar\gamma_{e}B_x \mathbf{\hat{S}_x} = \lambda_+\ket{\lambda_+}\bra{\lambda_+}+ \lambda_-\ket{\lambda_-}\bra{\lambda_-}+D\ket{-}\bra{-}
\end{equation}

where the eigenvalues are D and $\lambda_\pm=(D\pm\sqrt{D^2+(2\gamma_eB_x)^2})/2$. For the eigenstates we first define notation $\ket{\pm} = (\ket{m_s=1}\pm\ket{m_s-1})/\sqrt{2}$. The diagonal basis then consists of the three states $\ket{-}$, $\ket{\lambda_+}=\cos(\theta)\ket{+}+\sin(\theta)\ket{0}$ and $\ket{\lambda_-}=-\sin(\theta)\ket{+}+\cos(\theta)\ket{0}$ with $\tan(2\theta)=2\gamma_eB_x/D$.

Assuming $\gamma_e B_x\ll D$ then $\cos(\theta)\approx 1$ and $\sin(\theta)\approx0$:

\begin{equation}
    \hat{H}_{NV} = \lambda_+\ket{+}\bra{+} + \lambda_-\ket{0}\bra{0}+D\ket{-}\bra{-}
\end{equation}

The lowest energy state is $\ket{0}$ with energy $\lambda_-$, and we redefine the Hamiltonian with this state as the reference state that has zero energy.

\begin{equation}
    \hat{H}_{NV} = (\lambda_+-\lambda_-)\ket{+}\bra{+} + (D-\lambda_-)\ket{-}\bra{-}
\end{equation}

We can express the driving terms using the eigenstates $\ket{\pm}$ states, which gives $\mathbf{\hat{S}_y}=i(\ket{0}\bra{-}-\ket{-}\bra{0})$ and $\mathbf{\hat{S}_z}=\ket{+}\bra{-}+\ket{-}\bra{+}$. Substituting these expressions back into the full Hamiltonian including the driving terms, we obtain,

\begin{equation}
    H =(\lambda_+-\lambda_-)\ket{+}\bra{+} + (D-\lambda_-)\ket{-}\bra{-} + i\hbar\gamma_{e}B_{y}\cos(\omega_{y}t)(\ket{0}\bra{-}-\ket{-}\bra{0})+ \hbar\gamma_{e}B_{z}\cos(\omega_{z}t) (\ket{+}\bra{-}+\ket{-}\bra{+})
\end{equation}

We now switch the notation to match the Hamiltonian in the main text, where the operators $\hat{\sigma}_{\alpha\beta} = \ket{\alpha} \bra{\beta}$ describe the transition from $\ket{\beta}$ to $\ket{\alpha}$. The states are relabeled as follows $\ket{1}=\ket{0}$, $\ket{2}=\ket{-}$ and $\ket{3}=\ket{+}$. The frequencies of the states are defined as $\omega_{13}=\lambda_+-\lambda_-=\sqrt{D^2+(2\gamma_eB_x)^2}$ and $\omega_{12}=D-\lambda_- = (D+\sqrt{D^2+(2\gamma_eB_x)^2})/2$. We also introduce the resonator operators, with the resonator coupling to the transition between states $\ket{1}$ and $\ket{3}$, as well as the summation notation, since we are considering an ensemble of spins. Then the total Hamiltonian becomes,

\begin{equation}
H = \hbar \omega_{x} \hat{a}^{\dagger}\hat{a} +  \hbar \sum_{k=1}^{N} (\omega^{(k)}_{12}  \hat{\sigma}^{(k)}_{22} + \omega^{(k)}_{13}  \hat{\sigma}^{(k)}_{33}) \\+ \hbar \sum_{k=1}^{N} g_x(\hat{a}^{\dagger} \hat{\sigma}^{(k)}_{13}+ \hat{a}  \hat{\sigma}^{(k)}_{31})+i\hbar\gamma_{e}B_{y}\cos(\omega_{y}t) \sum_{k=1}^{N}( \hat{\sigma}^{(k)}_{12} -\hat{\sigma}^{(k)}_{21})\\
+\hbar\gamma_{e}B_{z}\cos(\omega_{z}t)\sum_{k=1}^{N}( 
\hat{\sigma}^{(k)}_{23} + \hat{\sigma}^{(k)}_{32})
\end{equation}
 
Finally we go into a rotating frame using a unitary operator $U=\exp(iH_0t/\hbar)$ with $H_0=\hbar\omega_{y}\sum_{k=1}^{N}\hat{\sigma}^{(k)}_{22}+\hbar(\omega_y + \omega_z)(\sum_{k=1}^{N}\hat{\sigma}^{(k)}_{33}+\hat{a}^{\dagger}\hat{a})$ and using the rotating-wave approximation to eliminate fast rotating terms, we arrive at the time independent Hamiltonian in the main text:

\begin{equation}
H = \hbar \Delta_{x} \hat{a}^{\dagger}\hat{a} + \hbar \sum_{k=1}^{N} (\Delta^{(k)}_{12}  \hat{\sigma}^{(k)}_{22} + (\Delta^{(k)}_{12} + \Delta^{(k)}_{23})  \hat{\sigma}^{(k)}_{33}) \\+ \hbar \sum_{k=1}^{N} g_x(\hat{a}^{\dagger} \hat{\sigma}^{(k)}_{13}+ \hat{a}  \hat{\sigma}^{(k)}_{31})+i\hbar\Omega_{y} \sum_{k=1}^{N}( \hat{\sigma}^{(k)}_{12} -\hat{\sigma}^{(k)}_{21})\\
+\hbar\Omega_{z}\sum_{k=1}^{N}( 
\hat{\sigma}^{(k)}_{23} + \hat{\sigma}^{(k)}_{32})
\end{equation}

where $\Delta_{x} =\omega_{x}-(\omega_y + \omega_z)$, $\Delta_{12}=\omega_{12}-\omega_{y}$, $\Delta_{23} = \omega_{23}- \omega_z$, $\Omega_{y}=\gamma_e B_{y}/2$ and $\Omega_{z}=\gamma_eB_{z}/2$.

\section{Verifying the masing phase transition}

We use the second-order correlation function $g^{(2)}(\tau)$ characterizes intensity correlations of the resonator field to verify the masing phase transition.
The $g^{(2)}(\tau)$ is defined as,
\begin{equation}
g^{(2)}(\tau)
=
\frac{
\left\langle
\hat{a}^{\dagger}(t)
\hat{a}^{\dagger}(t+\tau)
\hat{a}(t+\tau)
\hat{a}(t)
\right\rangle
}
{
\left\langle
\hat{a}^{\dagger}(t)\hat{a}(t)
\right\rangle
\left\langle
\hat{a}^{\dagger}(t+\tau)
\hat{a}(t+\tau)
\right\rangle
}.
\end{equation}
At zero delay $\tau=0$, the correlation function becomes
\begin{equation}
g^{(2)}(0)
=
\frac{
\left\langle
\hat{a}^{\dagger}
\hat{a}^{\dagger}
\hat{a}
\hat{a}
\right\rangle
}
{
\left|
\left\langle
\hat{a}^{\dagger}\hat{a}
\right\rangle
\right|^2
}.
\end{equation}

To express the fourth-order moment in terms of lower-order correlations, we employ the quantum cumulant expansion following \cite{kubo1962generalized,plankensteiner2022quantumcumulants}. 
For four operators, the fourth-order cumulant is \cite{kubo1962generalized,plankensteiner2022quantumcumulants},
\begin{equation}
\begin{split}
\langle ABCD\rangle_c
&=
\langle ABCD\rangle
-\Big[
\langle A\rangle\langle BCD\rangle
+\langle B\rangle\langle ACD\rangle
+\langle C\rangle\langle ABD\rangle
+\langle D\rangle\langle ABC\rangle
\Big]
\\
&\quad
-\Big[
\langle AB\rangle\langle CD\rangle
+\langle AC\rangle\langle BD\rangle
+\langle AD\rangle\langle BC\rangle
\Big]
+2\Big[
\langle A\rangle\langle B\rangle\langle CD\rangle
+\langle A\rangle\langle C\rangle\langle BD\rangle
+\langle A\rangle\langle D\rangle\langle BC\rangle
\\
&\quad
+\langle AB\rangle\langle C\rangle\langle D\rangle
+\langle AC\rangle\langle B\rangle\langle D\rangle
+\langle AD\rangle\langle B\rangle\langle C\rangle
\Big]
-6\langle A\rangle\langle B\rangle
\langle C\rangle\langle D\rangle .
\end{split}
\end{equation}
Similarly, third-order cumulants can also be expressed.
We make the assumption that the third-order cumulants and fourth-order cumulants are zero to simplify the $g^{(2)}(0)$. Then $\langle BCD\rangle_c=0,\langle ACD\rangle_c=0, \langle ABD\rangle_c=0,\langle ABC\rangle_c=0$, and $\langle ABCD\rangle_c=0$. This gives,
\begin{align}
\langle BCD\rangle
&=
\langle BC\rangle\langle D\rangle
+\langle BD\rangle\langle C\rangle
+\langle CD\rangle\langle B\rangle
-2\langle B\rangle\langle C\rangle\langle D\rangle ,
\\
\langle ACD\rangle
&=
\langle AC\rangle\langle D\rangle
+\langle AD\rangle\langle C\rangle
+\langle CD\rangle\langle A\rangle
-2\langle A\rangle\langle C\rangle\langle D\rangle ,
\\
\langle ABD\rangle
&=
\langle AB\rangle\langle D\rangle
+\langle AD\rangle\langle B\rangle
+\langle BD\rangle\langle A\rangle
-2\langle A\rangle\langle B\rangle\langle D\rangle ,
\\
\langle ABC\rangle
&=
\langle AB\rangle\langle C\rangle
+\langle AC\rangle\langle B\rangle
+\langle BC\rangle\langle A\rangle
-2\langle A\rangle\langle B\rangle\langle C\rangle,
\end{align}
and
\begin{equation}
\langle ABCD\rangle
=
\langle AB\rangle\langle CD\rangle
+\langle AC\rangle\langle BD\rangle
+\langle AD\rangle\langle BC\rangle
-2\langle A\rangle
\langle B\rangle
\langle C\rangle
\langle D\rangle .
\end{equation}

Setting $A=a^\dagger,B=a^\dagger,C=a,D=a$ the fourth-order moment becomes:

\begin{equation}
\langle a^\dagger a^\dagger a a\rangle
=
|\langle aa\rangle|^2
+
2|\langle a^\dagger a\rangle|^2
-
2|\langle a\rangle|^4 .
\end{equation}

Consequently, the zero-delay second-order correlation function is

\begin{equation}
g^{(2)}(0)
=
\frac{
|\langle aa\rangle|^2
+
2|\langle a^\dagger a\rangle|^2
-
2|\langle a\rangle|^4
}
{
|\langle a^\dagger a\rangle|^2
}.
\end{equation}

The above expression depends only on first-order and second-order terms and is valid for gaussian states, including vacuum states, thermal states, and coherent states, which are fully characterized by first and second-order moments of the field operator  \cite{RevModPhys.84.621}.

\begin{figure}[H]
    \centering
    \includegraphics[width=0.4\linewidth]{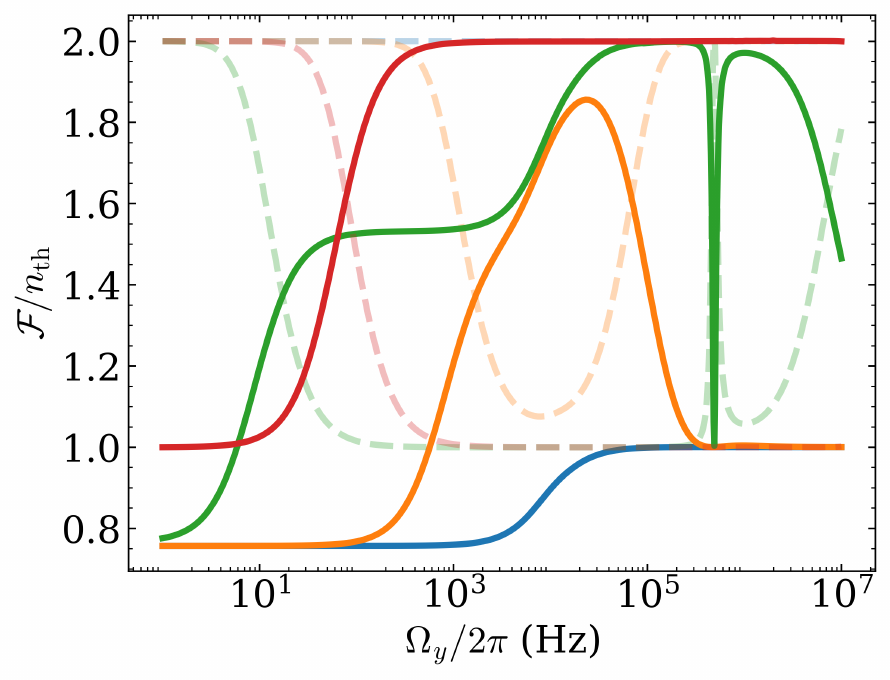}
    \caption{Fano factor $\mathcal{F}/n_{\mathrm{th}}$ plotted plotted as a function of the driving strength $\Omega_{y}$ for different values of $\Omega_{z}/2\pi=10$~Hz (blue), 10$^3$~Hz (orange), 10$^5$~Hz (green), 10$^7$~Hz (red). $g^{(2)}(0)$ is plotted with dashed-lines to indicate where the phase transitions for each curve.}
    \label{fig:s1}
\end{figure}

The masing phase transition can also be verified using other parameters that characterize the photon statistics, such as the Fano factor, $\mathcal{F}=\langle\hat{a}^\dagger\hat{a}\hat{a}^\dagger\hat{a}\rangle/\langle\hat{a}^\dagger\hat{a}\rangle$, which characterizes the fluctuation in the photon number \cite{werren2022quantum}.
Reducing the fourth-order terms similar to $g^{(2)}(0)$ in to first and second order terms, $\mathcal{F}$ can be expressed as,
\begin{equation}
\mathcal{F} = 1+\langle \hat{a}^\dagger\hat{a} \rangle(g^{(2)}(0)-1)
\label{eq:fano}
\end{equation}
We plot $\mathcal{F}$ in Fig.~\ref{fig:s1} using the same color coding in Fig.~2(d) of the main text. Conventionally, we expect a peak in the $\mathcal{F}$ at the midpoint of the phase transition \cite{PhysRevA.50.4318}, however, we observe that $\mathcal{F}$ increases or decreases along the phase transition with no distinguishable peak.
Although the absence of peaks for $\mathcal{F}$ does not rule out the phase transition, an accurate estimation of $\mathcal{F}$ may require more detailed analysis \cite{saldutti2024onset} 

In Fig.~\ref{fig:s1} for the cases of $\Omega_z/2\pi=10$~Hz, $1$~kHz and $100$~kHz, and for weak driving with $\Omega_y$, the factor $\mathcal{F}/n_{\mathrm{th}}<1$. In this region, since $g^{(2)}(0)=2$, we expect $\mathcal{F}/n_{\mathrm{th}}\approx1$, however, the steady state photon number in this region goes below the thermal photon occupation number ($n_x<n_{\mathrm{th}}$). We attribute this reduction in photon number to resonator mode cooling provided by the spins and the driving fields.
 
The linewidth narrowing across the phase transition can also be an indicator in a typical masing phase transition. 
The accurate estimation of the linewidth requires careful consideration of the noises in the system, which results in spectral feautures with finite linewidths.
In our case, we expect the requirement of a detailed consideration of noises due to the driving fields, spin frequency distribution due to inhomogeneous broadening, resonator fluctuations, etc., and this lies beyond the scope of the present work.

\section{Table of rates for Lindblad operators}

\begin{table*}[h]
\caption{Dissipative processes included in the seven-level NV model. The transition operators are defined as
$\hat{\sigma}_{ij}^{(k)}=|i\rangle_k\langle j|$. The pure dephasing, longitudinal relaxation, and longitudinal excitation operators are expressed in the rotated ground-state basis employed in the model.}
\label{tab:dissipation_rates}
\centering

\begin{tabular}{lll}
\hline\hline
Process & Jump operator & Rate \\
\hline

Optical decay $4\rightarrow1$
& $\sqrt{\gamma_{14}}\,\hat{\sigma}_{14}$
& $\gamma_{14}/2\pi = 66.16~\mathrm{MHz}$\cite{gupta2016efficient} \\

Optical decay $5\rightarrow2$
& $\sqrt{\gamma_{25}}\,\hat{\sigma}_{25}$
& $\gamma_{25}/2\pi = 66.16~\mathrm{MHz}$\cite{gupta2016efficient} \\

Optical decay $6\rightarrow3$
& $\sqrt{\gamma_{36}}\,\hat{\sigma}_{36}$
& $\gamma_{36}/2\pi = 66.16~\mathrm{MHz}$\cite{gupta2016efficient} \\

ISC $5\rightarrow7$
& $\sqrt{\gamma_{75}}\,\hat{\sigma}_{75}$
& $\gamma_{75}/2\pi = 91.8~\mathrm{MHz}$\cite{gupta2016efficient} \\

ISC $6\rightarrow7$
& $\sqrt{\gamma_{76}}\,\hat{\sigma}_{76}$
& $\gamma_{76}/2\pi = 91.8~\mathrm{MHz}$\cite{gupta2016efficient} \\

ISC $4\rightarrow7$
& $\sqrt{\gamma_{74}}\,\hat{\sigma}_{74}$
& $\gamma_{74}/2\pi = 11.1~\mathrm{MHz}$\cite{gupta2016efficient} \\

Singlet decay $7\rightarrow2$
& $\sqrt{\gamma_{27}}\,\hat{\sigma}_{27}$
& $\gamma_{27}/2\pi = 2.04~\mathrm{MHz}$\cite{gupta2016efficient} \\

Singlet decay $7\rightarrow3$
& $\sqrt{\gamma_{37}}\,\hat{\sigma}_{37}$
& $\gamma_{37}/2\pi = 2.04~\mathrm{MHz}$\cite{gupta2016efficient} \\

Singlet decay $7\rightarrow1$
& $\sqrt{\gamma_{17}}\,\hat{\sigma}_{17}$
& $\gamma_{17}/2\pi = 4.87~\mathrm{MHz}$\cite{gupta2016efficient} \\

Resonator loss
& $\sqrt{\kappa_x}\,\hat{a}$
& $\kappa_x/2\pi = 130~\mathrm{kHz}$\cite{wang2025cavity} \\

Pure dephasing
& $L_{p\pm}$
& $\Gamma/2\pi = 330~\mathrm{kHz}$\cite{wang2025cavity} \\

Longitudinal relaxation
& $L_{r\pm}$
& $\gamma_l/2\pi = 200~\mathrm{Hz}$\cite{jarmola2012temperature} \\

Longitudinal excitation
& $L_{e\pm}$
& $\gamma_l/2\pi = 200~\mathrm{Hz}$\cite{jarmola2012temperature} \\

\hline\hline

\end{tabular}
\end{table*}

\begin{align}
L_{p\pm}
&=
\sqrt{\frac{\Gamma}{2}}
\left(
\hat{\sigma}^{(k)}_{22}
+
\hat{\sigma}^{(k)}_{33}
\pm
\hat{\sigma}^{(k)}_{23}
\pm
\hat{\sigma}^{(k)}_{32}
\right),
\\[4pt]
L_{r\pm}
&=
\sqrt{\frac{\gamma_l}{2}}
\left(
\hat{\sigma}^{(k)}_{13}
\pm
\hat{\sigma}^{(k)}_{12}
\right),
\\[4pt]
L_{e\pm}
&=
\sqrt{\frac{\gamma_l}{2}}
\left(
\hat{\sigma}^{(k)}_{31}
\pm
\hat{\sigma}^{(k)}_{21}
\right).
\end{align}

\section{Effects of magnetic field detuning under strong driving field}
\setcounter{equation}{0}
\setcounter{figure}{0}

\begin{figure}[H]
    \centering
    \includegraphics[width=0.4\linewidth]{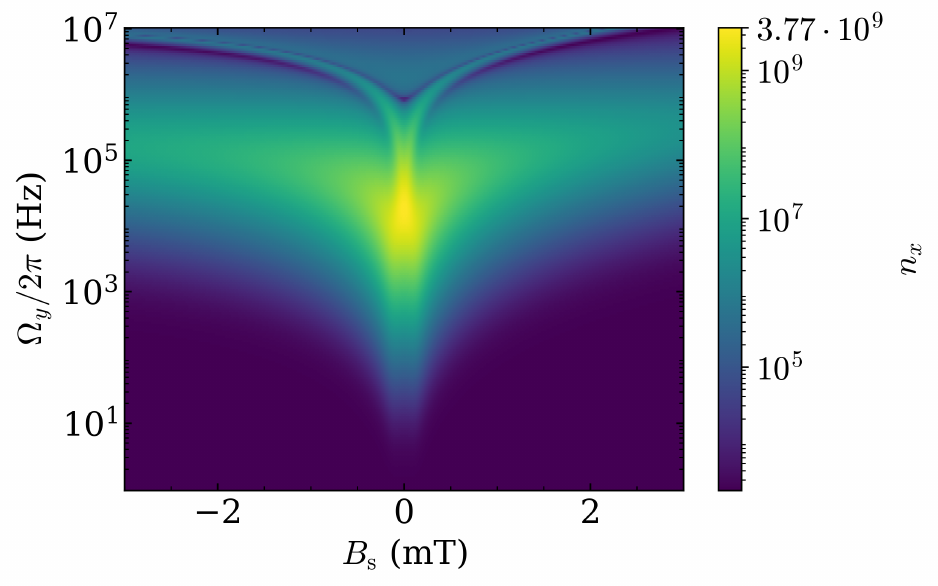}
    \caption{Photon number as a function of microwave driving strength $\Omega_y$ and magnetic field offset parameter $B_{s}$. In the weak driving regime ($\Omega_y < \Gamma$), the response is dominated by the bare spin transition and exhibits two peaks with very small splitting. At strong driving ($\Omega_y > \Gamma$), coherent driving produces dressed-state splitting, leading to two distinct gain resonances whose separation increases with $\Omega_y$.}
    \label{fig:s2}
\end{figure}

To investigate the response under strong microwave driving, we numerically calculate the $n_{x}$ as a function of the driving strength $\Omega_y$ and magnetic field offset $B_s$. The results are shown in Fig.~\ref{fig:s2} for fixed $\Omega_z/2\pi=500$~kHz and $\Lambda/2\pi = 500$~Hz.
In the weak driving regime, where $\Omega_y$ remains smaller than the intrinsic spin dephasing rate $\Gamma$, the response is primarily governed by the spin transition frequency. 
Consequently, the gain exhibits a maximum close to zero magnetic detuning, with the response gradually degrading as the spin transition is shifted away from resonance. 
Although we expect a single resonance in this region, we observe two small peaks likely formed due to the presence of the other drive $\Omega_z$.
Near the region of maximum gain, the linewidth also narrows.
As the driving strength is increased beyond the spin dephasing rate ($\Omega_y > \Gamma$), the system enters the strong-driving regime, where the coherent driving field dominates over dissipative processes. In this regime, the microwave field produces a splitting of the dressed-state energy levels. This splitting becomes more pronounced with increasing $\Omega_y$, leading to two distinct resonant features in the gain spectrum.
The magnetic field detuning $\delta B$ shifts the spin transition frequency relative to the dressed-state resonances. As a result, the photon number response evolves from a nearly single resonance at weak driving to a clearly resolved double-peaked structure at strong driving. 

\section{Julia code for the seven-level driven model}

The following Julia implementation generates the first-order mean-field equations and provides an example of a numerical solution. The second-order equations can be obtained by setting \texttt{order = 2} and redefining the ops array as \texttt{ops = [$\hat{a}^\dagger \hat{a}$]}.

\begin{lstlisting}[language=Julia]
using QuantumCumulants
using ModelingToolkit
using DifferentialEquations
# number of levels for model
l = 7
# define spin parameters
N, Δ12, Δ23, Ωy, Ωz, γl, Γ, Λ, γ14, γ25, γ36, γ74, γ75, γ76, γ17, γ27, γ37 
= cnumbers("N Δ_{12} Δ_{23} Ω_y Ω_z γ_l Γ Λ γ_{14} γ_{25} γ_{36} γ_{74} γ_{75} γ_{76} 
γ_{17} γ_{27} γ_{37}")
# define resonator parameters
Δx, κx = cnumbers("Δ_x κ_x")
# define Hilbert space
hc = FockSpace(:resonator)
hs = NLevelSpace(:spin, l)
h = hc ⊗ hs
# resonator operator
@qnumbers a::Destroy(h)
# spin operators
σ(α,β,i) = IndexedOperator(Transition(h,:σ,α,β),i)
# single spin coupling
gx(i) = IndexedVariable(:gx,i)
# summation index
i = Index(h,:i,N,hs)
# Hamiltonian
Hs = Δ12*Σ(σ(2,2,i),i) + (Δ12+Δ23)*Σ(σ(3,3,i),i)
Hx = Δx*a'*a + Σ(gx(i)*(a'*σ(1,3,i)+a*σ(3,1,i)),i)
Hy = 1im*Ωy*Σ(σ(1,2,i)-σ(2,1,i),i)
Hz = Ωz*Σ(σ(3,2,i)+σ(2,3,i),i)
H = Hs + Hx + Hy + Hz
# generate mean-field equations
ops=[a]
eqs =meanfield(ops,H,jumpops;rates=rates,order=1)
eqscompleted = complete(eqs)
eqssc = scale(eqscompleted)
@named sys = ODESystem(eqssc)
# parameter values
B = 5e-3
ge = 2*pi*28e9
D = 2*pi*2.87e9
ω13 = sqrt(D^2 + (2*ge*B)^2)
ω12 = (D + sqrt(D^2 + (2*ge*B)^2))/2
# set driving frequencies to spin transitions
ωy = ω12
ωz = ω13 - ω12
# set resonator frequency to spin transition
ωx = ω13
# detunings
Δ23val = ω13 - ω12 - ωz  
Δ12val = ω12 - ωy
Δxval = ωx - (ωz + ωy)
# driving strengths
Ωyval = 2*pi*33.5e3
Ωzval = 2*pi*628e3
# optical pump rate
Λval = 2*pi*500
# number of spins
Nval = 1.1e14
# single spin coupling
gxval = 2*pi*18e-3
# relaxation rates
γlval = 2*pi*200
# dephasing rate
Γval = 2*pi*330e3
# resonator loss
κxval = 2*pi*130e3
# decay rates
γ14val = 2*pi*66.16e6
γ25val = 2*pi*66.16e6
γ36val = 2*pi*66.16e6
γ74val = 2*pi*11.1e6
γ75val = 2*pi*91.8e6
γ76val = 2*pi*91.8e6
γ17val = 2*pi*4.87e6
γ27val = 2*pi*2.04e6
γ37val = 2*pi*2.04e6
# map parameters to values
ps = (Δ12, Δ23, Δx, Ωz, Ωy, gx(1), N, κx, γl, Γ, Λ, γ14, γ25, γ36, γ74, γ75, γ76, γ17, γ27, γ37)
p0 = (Δ12val, Δ23val, Δxval, Ωzval, Ωyval, gxval, Nval, κxval, γlval,Γval,
    Λval, γ14val, γ25val, γ36val, γ74val, γ75val, γ76val, γ17val, γ27val, γ37val)
# example dynamic solution
tf = 5e-4
u0 = zeros(ComplexF64, length(eqssc))
dict = merge(Dict(unknowns(sys) .=> u0),Dict(ps .=> p0))
prob = ODEProblem(sys, dict, (0.0, tf))
sol = solve(prob, BS3(),maxiters=10000000,reltol=1e-8,abstol=1e-8);
\end{lstlisting}

\begin{figure}[H]
    \centering
    \includegraphics[width=0.5\linewidth]{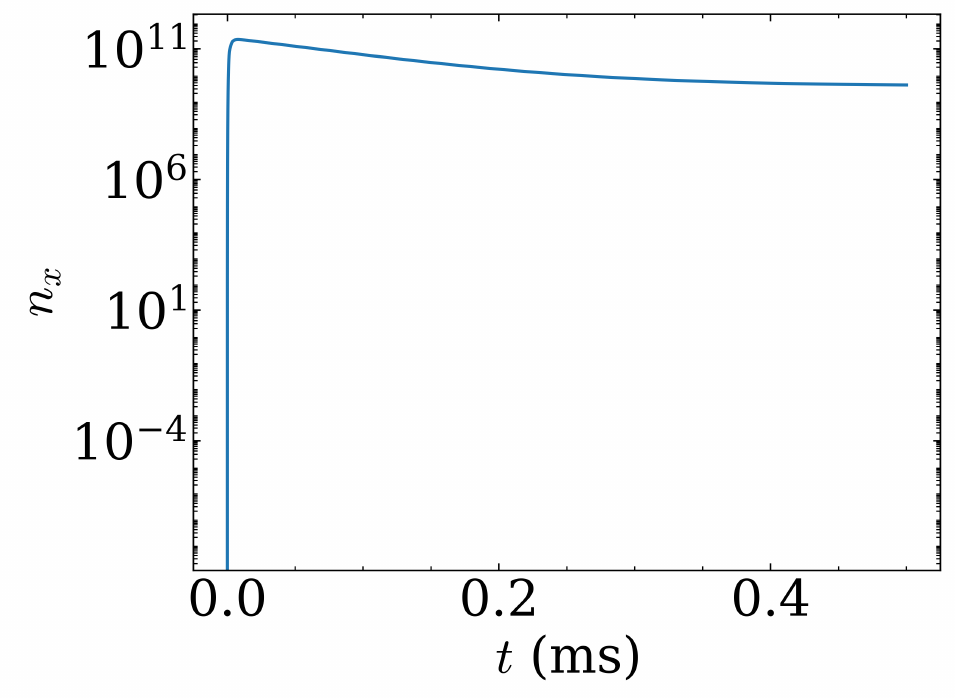}
    \caption{Example transient solution for the specific case of the best sensitivity in Fig.~3(a)}. We set the initial condition for $\langle a \rangle = 0$, and even then, there is a resultant $n_{x}$ which suggests our system does not hold the phase-invariant condition. Please also note that there is no oscillation in $n_{x}$ obtained from our model, before reaching the steady state, unlike in the conventional population inversion.
    \label{fig:s3}
\end{figure}

Figure \ref{fig:s3} is an example output of the transient solution from the code above, showing the photon number as a function of time.

\section{Photon number reduction in Fig.~2(a) of main text}

\begin{figure}[H]
    \centering
    \includegraphics[width=0.3\linewidth]{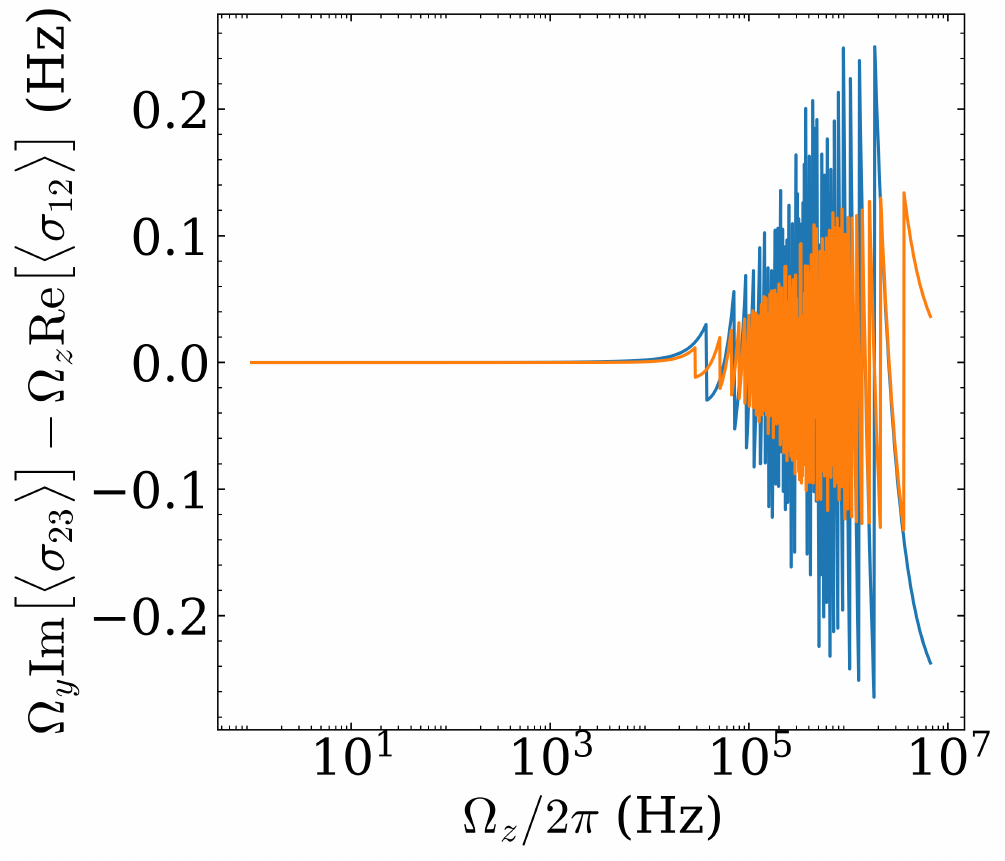}
    \caption{Quantity $\Omega_y \mathrm{Im}[\langle \sigma_{23} \rangle]-\Omega_z \mathrm{Re}[\langle \sigma_{12} \rangle]$ evaluated at the photon-number minimum indicated by the red dashed line in Fig.~2(a). At this point, the coherences generated by the two driving fields nearly cancel, causing the last two terms of the gain [Eq.~(12)] to approach zero. The blue and orange curves were obtained using mesh sizes of $1000\times1000$ and $2000\times2000$ simulated points, respectively. The improved resolution demonstrates convergence toward zero, suggesting the cancellation of the coherence contributions.}
    \label{fig:s4}
\end{figure}

To elucidate the reduction in the photon number observed in Fig.~2(a) of the main text (red dashed line), we plot in Fig.~\ref{fig:s4} the quantity $\Omega_y \mathrm{Im}[\langle \sigma_{23} \rangle] - \Omega_z \mathrm{Re}[\langle \sigma_{12} \rangle]$ as a function of $\Omega_z$, evaluated at the photon-number minimum shown in Fig.~2(a). This quantity contributes to the microwave gain $G$ through the last two terms of Eq.~(12) (see main text). For zero detuning, $\tilde{\Delta}=0$, the $\langle a\rangle$ obtained from our numerical solution remains purely real throughout this regime. Consequently, the last two terms of the gain expression are proportional to $\left(\Omega_y \mathrm{Im}[\langle \sigma_{23} \rangle] - \Omega_z \mathrm{Re}[\langle \sigma_{12} \rangle]\right)$. The reduction in photon number therefore arises from the cancellation of the coherence contributions induced by the two driving fields, which suppresses the corresponding contribution to the microwave gain.

\section{Magnetic-field Sensitivity Optimization}

\begin{figure}[H]
    \centering
    \includegraphics[width=0.7\linewidth]{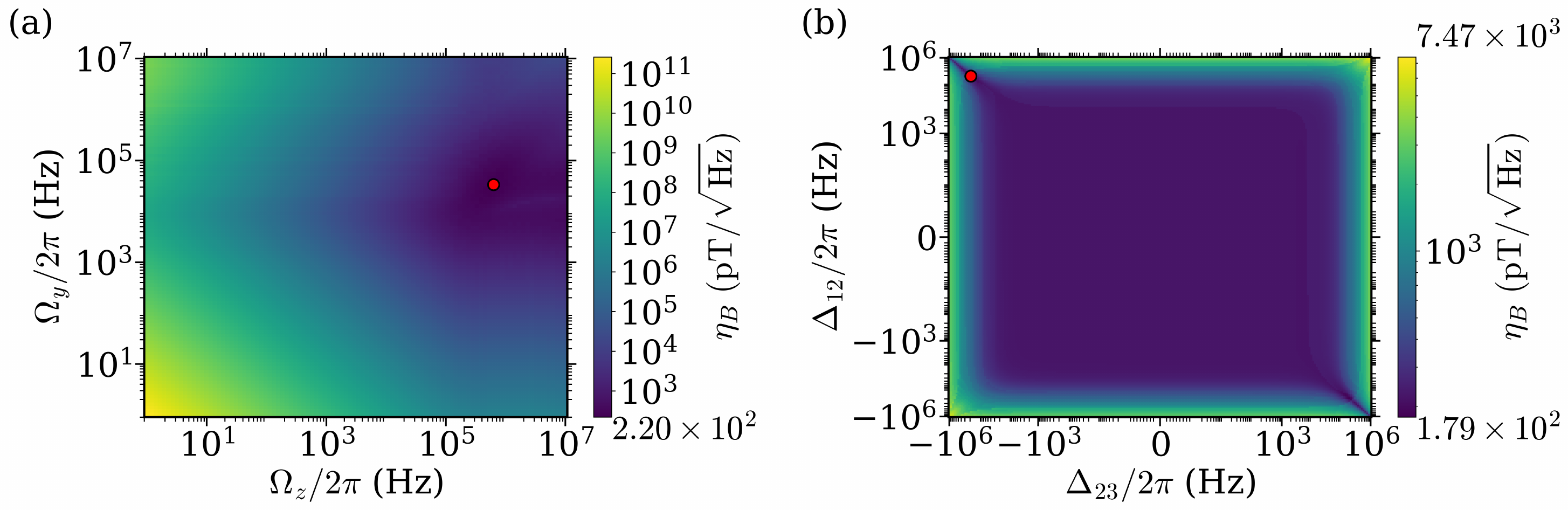}
    \caption{(a) Magnetic field sensitivity $\eta_B$ as a function of driving field strengths $\Omega_{y}$ and $\Omega_{z}$ for $\Lambda/2\pi=500$~Hz. The red dot indicates the minimum sensitivity $220~\mathrm{pT/\sqrt{Hz}}$ attainable with resonant driving fields ($\Delta_{12}=\Delta_{23}=0$) at $\Omega_y/2\pi = 33.5$~kHz and $\Omega_z/2\pi=628$~kHz. (b) Magnetic field sensitivity $\eta_B$ as a function of detunings $\Delta_{12}$ and $\Delta_{23}$ for fixed driving strengths   $\Omega_{y}/2\pi = 33.5$~kHz, $\Omega_{z}/2\pi = 628$~kHz and pump rate $\Lambda/2\pi=500$~Hz. At detunings of $\Delta_{12}/2\pi=187$~kHz, $\Delta_{23}/2\pi=-187$~kHz (see red dot), the sensitivity reduces further to $179~\mathrm{pT/\sqrt{Hz}}$ }
    \label{fig:s5}
\end{figure}

To determine the optimal magnetic field sensitivity, we numerically optimized the driving parameters of the sensing scheme as follows.
We first considered resonant driving ($\Delta_{12}=\Delta_{23}=0$) and calculated the sensitivity over a wide range of driving strengths $\Omega_y$ and $\Omega_z$, while keeping the optical pump rate fixed at $\Lambda/2\pi=500$~Hz. As shown in Fig.~\ref{fig:s5}(a), the optimal resonant operating point occurs at $\Omega_y/2\pi=33.5$~kHz and $\Omega_z/2\pi=628$~kHz, yielding a minimum sensitivity of $220~\mathrm{pT/\sqrt{Hz}}$. 
We then fixed these optimized driving strengths and varied the microwave detunings $\Delta_{12}$ and $\Delta_{23}$ by varying $\omega_y$ and $\omega_z$, respectively. Figure~\ref{fig:s5}(b) shows that introducing asymmetric detunings further enhances the sensitivity, with the optimum found at $\Delta_{12}/2\pi=187$~kHz and $\Delta_{23}/2\pi=-187$~kHz, reducing the sensitivity to $179~\mathrm{pT/\sqrt{Hz}}$. 

\begin{figure}[H]
    \centering
    \includegraphics[width=0.5\linewidth]{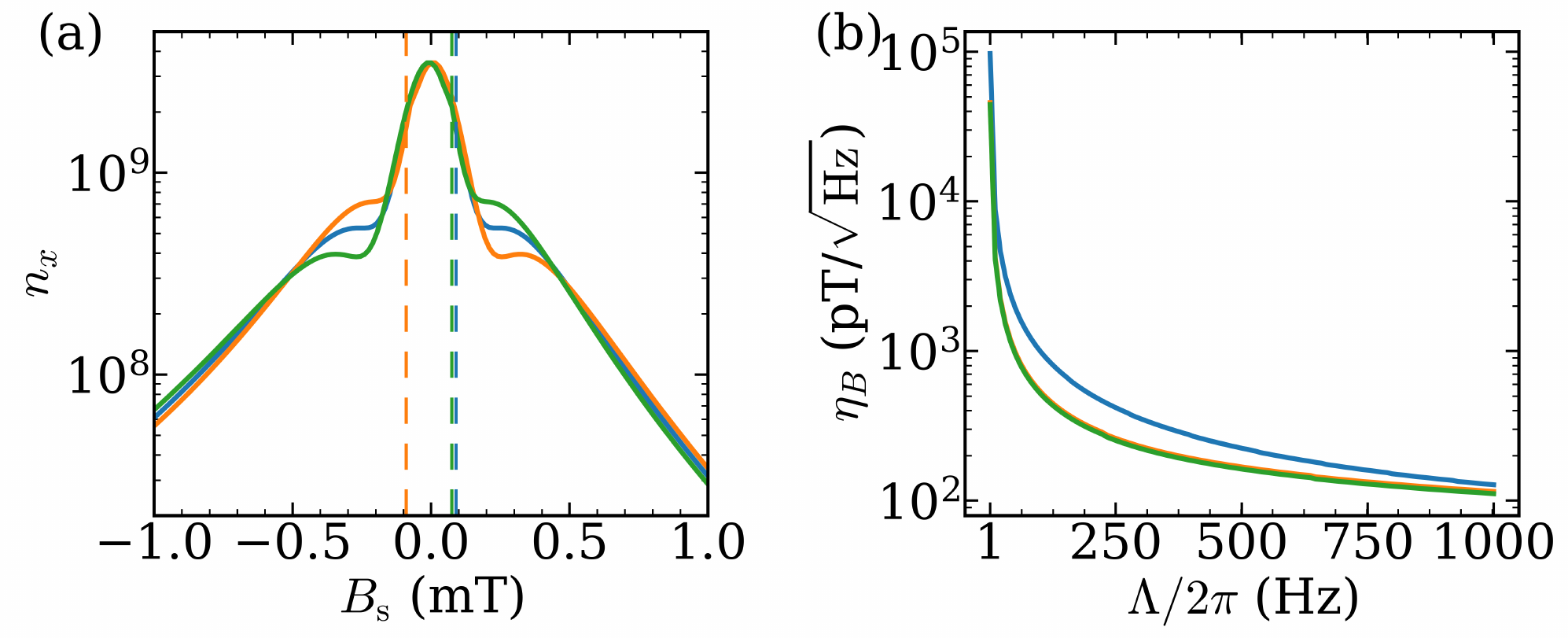}
    \caption{(a) Photon number $n_x$ as a function of $B_\mathrm{s}$ with $\Omega_y/2\pi = 33.5$~kHz, $\Omega_z/2\pi=628$~kHz, $\Lambda/2\pi = 500$~Hz with detunings $\Delta_{12}/2\pi=\Delta_{23}/2\pi=0$~(blue) and detuning combinations $\Delta_{12}/2\pi=-187$~kHz, $\Delta_{23}/2\pi=187$~kHz~(orange) and $\Delta_{12}/2\pi=187$~kHz, $\Delta_{23}/2\pi=-187$~kHz~(green) which is the point where minimum sensitivity occurs based on optimization of parameters in Fig.~\ref{fig:s5}. (b) Magnetic sensitivity as a function of $\Lambda$ for each case in (a).}
    \label{fig:s6}
\end{figure}

Figure~\ref{fig:s6}(a) compares the profile of $n_x$ as a function of $B_\mathrm{s}$ with the detuning combinations at the minimum-sensitivity operating points identified in Fig.~\ref{fig:s6}(b) with the symmetric case, $\Delta_{12}/2\pi=\Delta_{23}/2\pi=0$. For finite drive detuning, the lineshape becomes asymmetric about $B_\mathrm{s}=0$. This asymmetry originates from the detuning-dependent spin-resonator response, which no longer remains invariant under the transformation $B_\mathrm{s}\rightarrow -B_\mathrm{s}$.
To demonstrate this, we define the detunings
\begin{align}
\Delta_x &= \omega_x-(\omega_y+\omega_z)\\
\Delta_{12}(B_{\mathrm{tot}}) &= \omega_{12}(B_{\mathrm{tot}})-\omega_y\\
\Delta_{13}(B_{\mathrm{tot}}) &= \omega_{13}(B_{\mathrm{tot}})-(\omega_y+\omega_z)
\end{align}
where $B_{\mathrm{tot}}$ denotes the total magnetic field along the $x$-axis. The transition frequencies are
\begin{align}
\omega_{12}(B_{\mathrm{tot}})&=\frac{D+\sqrt{D^2+(2\gamma_e B_{\mathrm{tot}})^2}}{2}\\
\omega_{13}(B_{\mathrm{tot}})&=\sqrt{D^2+(2\gamma_e B_{\mathrm{tot}})^2}
\end{align}

In the weak-field limit, $\gamma_e B_{\mathrm{tot}} \ll D$, these reduce to
\begin{align}
\omega_{12}(B_{\mathrm{tot}})&\approx D+\frac{(\gamma_e B_{\mathrm{tot}})^2}{D}\\
\omega_{13}(B_{\mathrm{tot}})&\approx D+\frac{2(\gamma_e B_{\mathrm{tot}})^2}{D}
\end{align}

We decompose the magnetic field as
\begin{equation}
B_{\mathrm{tot}} = B_x + B_{\mathrm{s}},
\end{equation}
where $B_x = 5~\mathrm{mT}$ is the static bias field and $B_{\mathrm{s}} \ll B_x$. Expanding to first-order in $B_{\mathrm{s}}$ gives

\begin{align}
\omega_{12}(B_\mathrm{s})&\approx D+\frac{(\gamma_eB_x)^2}{D}
+\frac{2\gamma_e^2B_xB_\mathrm{s}}{D}\\
\omega_{13}(B_\mathrm{s})&\approx D+\frac{2(\gamma_eB_x)^2}{D}
+\frac{4\gamma_e^2B_xB_\mathrm{s}}{D}
\end{align}

Next, we allow the drive frequencies to vary according to
\begin{align}
\omega_y&=\omega_{y0}+\delta\omega_y\\
\omega_z&=\omega_{z0}+\delta\omega_z
\end{align}
where $\omega_x=\omega_{13}(B_\mathrm{s}=0)$, $\omega_{y0}=\omega_{12}(B_\mathrm{s}=0)$, and $\omega_{z0}=\omega_{23}(B_\mathrm{s}=0)$. Since the resonator and driving fields are resonant at $B_\mathrm{s}=0$, each field addresses its corresponding transition resonantly. At the optimal operating point identified in Fig.~\ref{fig:s5}, the drive detunings satisfy $\delta\omega_y=-\delta\omega_z$.

The resulting detunings are therefore
\begin{align}
\Delta_x(B_\mathrm{s})&=0\\
\Delta_{12}(B_\mathrm{s})&=\frac{2\gamma_e^2B_xB_\mathrm{s}}{D}-\delta\omega_y\\
\Delta_{13}(B_\mathrm{s})&=\frac{4\gamma_e^2B_xB_\mathrm{s}}{D}
\end{align}

In the absence of drive detuning ($\delta\omega_y = \delta\omega_z = 0$), the detunings are linear in $B_\mathrm{s}$ and the system is symmetric under $B_\mathrm{s} \rightarrow -B_\mathrm{s}$. Finite detuning introduces a constant offset that breaks this symmetry, leading to an asymmetric response in $B_\mathrm{s}$.

For finite detuning ($\delta\omega_y\neq0$), the origin of the asymmetry is evident under the transformation $B_\mathrm{s}\rightarrow -B_\mathrm{s}$:

\begin{align}
\Delta_{12}(-B_\mathrm{s})&=-\frac{2\gamma_e^2B_xB_\mathrm{s}}{D}-\delta\omega_y
\neq-\Delta_{12}(B_\mathrm{s})\\
\Delta_{13}(-B_\mathrm{s})&=-\Delta_{13}(B_\mathrm{s})
\end{align}
Thus, only $\Delta_{13}$ changes sign, whereas $\Delta_{12}$ retains the constant offset arising from the drive detuning. 

We also plot the sensitivity as a function of pump rate in Fig.~\ref{fig:s6}(b), showing enhanced sensitivity for $\Lambda\in[0,1000]$~Hz with the best sensitivities occurring when $\Delta_{12}/2\pi=187$~kHz, $\Delta_{23}/2\pi=-187$~kHz.

\begin{figure}[H]
    \centering
    \includegraphics[width=0.25\linewidth]{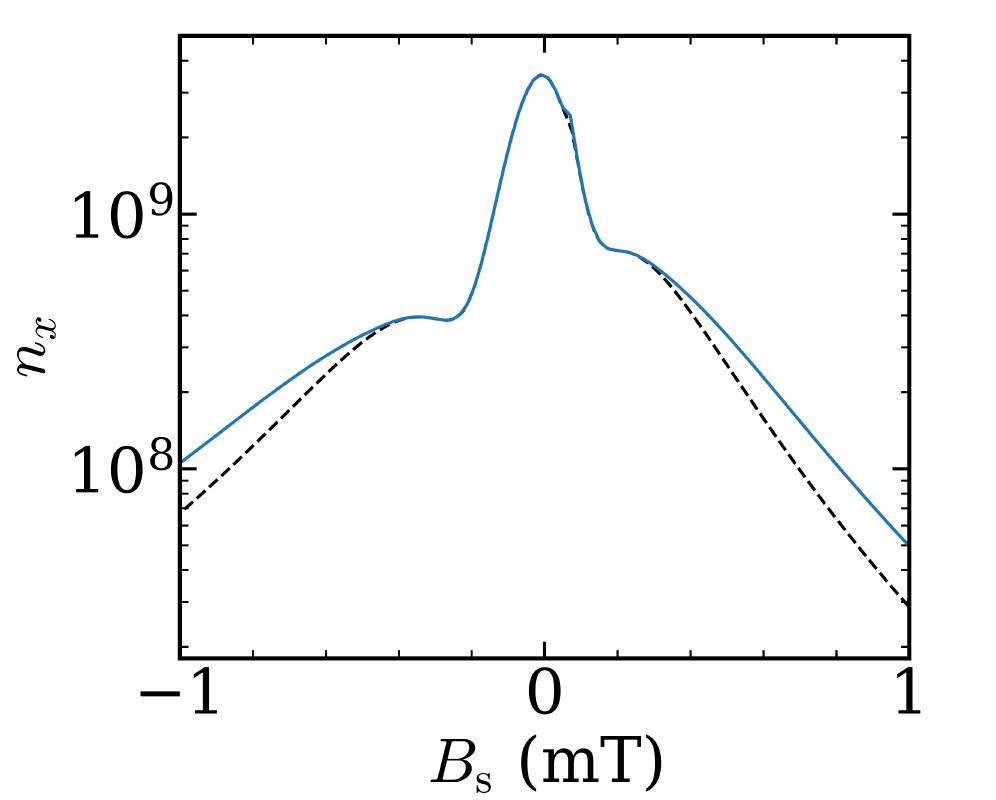}
    \caption{$n_x$ as a function of $B_\mathrm{s}$ calculated using second-order correlated equations (blue line) using the same parameters from Fig.~3(a) in the main text. For comparison, the dashed line represents the results calculated from the first-order equations. The calculated sensitivities for this case were 161 (179)~pT/$\sqrt{Hz}$ for the second (first) order equations.}
    \label{fig:s7}
\end{figure}

We additionally investigate the impact of higher-order spin correlations on the magnetometry. We compare the magnetic response calculated using the second-order correlated equations with that obtained from the first-order equations. 
For the second-order equations, a total of 75 equations were solved up to a time of 0.5~ms, and the final values of $n_x$ were used to calculate the sensitivity at each value of $B_\mathrm{s}$.
Figure~\ref{fig:s7} shows that the second-order treatment predicts a slightly steeper dependence of $n_x$ on the magnetic field $B_\mathrm{s}$, while the overall lineshape remains nearly unchanged. 
This enhanced slope translates directly into an improved magnetic field sensitivity, yielding $161~\mathrm{pT/\sqrt{Hz}}$ compared with $179~\mathrm{pT/\sqrt{Hz}}$ obtained from the first-order equations. 

\bibliography{references.bib}